\documentclass[12pt]{article}
\usepackage{epsfig}
\usepackage{amssymb,amsmath}
\newcommand{\lsim}{\,{\buildrel < \over {_\sim}}\,}
\newcommand{\gsim}{\,{\buildrel > \over {_\sim}}\,}

\newcommand{\comment}[1]{}

\begin{document}

\begin{titlepage}
\begin{flushright}
HIP-2007-10/TH\\
Roma1-1448/2007\\
hep-ph/0703104\\
\today
\end{flushright}
\vfill
\begin{centering}

{\bf A GLOBAL REANALYSIS OF NUCLEAR \\ PARTON DISTRIBUTION FUNCTIONS}

\vspace{0.5cm}
K.J. Eskola$^{\rm a,b,}$\footnote{kari.eskola@phys.jyu.fi},
V.J. Kolhinen$^{\rm a,b,}$\footnote{vesa.kolhinen@phys.jyu.fi},
H. Paukkunen$^{\rm a,b,}$\footnote{hannu.paukkunen@phys.jyu.fi} and 
C.A. Salgado$^{\rm c,}$\footnote{carlos.salgado@cern.ch}$^,$\footnote{Permanent address: Departamento 
de F\'\i sica de Part\'\i culas, Universidade de Santiago de Compostela, Spain}

\vspace{1cm}
{\em $^{\rm a}$Department of Physics,
P.O. Box 35, FI-40014 University of Jyv\"askyl\"a, Finland}
\vspace{0.3cm}

{\em $^{\rm b}$Helsinki Institute of Physics,
P.O. Box 64, FI-00014 University of Helsinki, Finland}
\vspace{0.3cm}

{\em $^{\rm c}$Dipartimento di Fisica, Universit\`a  di Roma ``La Sapienza'' and INFN, 
Roma, Italy}

\vspace{1cm} 
{\bf Abstract} \\ 
\end{centering}

\noindent
We determine the nuclear modifications of parton distribution functions
of bound protons at scales $Q^2\ge 1.69$~GeV$^2$ and momentum fractions 
$10^{-5}\le x\le 1$ in a global analysis which utilizes nuclear hard process data, 
sum rules and leading-order DGLAP scale evolution. The main improvements over 
our earlier work {\em EKS98} are the automated $\chi^2$ minimization, simplified and better 
controllable fit functions, and most importantly, the possibility for error estimates.
The resulting 16-parameter fit to the $N=514$ datapoints is good, $\chi^2/{\rm d.o.f}=0.82$. 
Within the error estimates obtained, the old {\em EKS98} parametrization is found 
to be fully consistent with the present analysis, with no essential difference in terms of $\chi^2$ either. We also determine separate uncertainty bands for the nuclear gluon and sea quark modifications in the 
large-$x$ region where they are not stringently constrained by the available data. Comparison with 
other global analyses is shown and uncertainties demonstrated. Finally, we show that
RHIC-BRAHMS data for inclusive hadron production in d+Au collisions lend support for a stronger gluon 
shadowing at $x<0.01$ and also that fairly large changes in the gluon modifications do not rapidly 
deteriorate the goodness of the overall fits, as long as the initial gluon modifications in the 
region $x\sim 0.02-0.04$ remain small.

\vfill
\end{titlepage}

\setcounter{footnote}{0}

\section{Introduction}

Universal, process-independent parton distribution functions (PDFs) of free and bound nucleons 
are a key element in the computational phenomenology of processes involving large virtualities $Q^2$ 
in hadronic and nuclear collisions. The free proton PDFs are nowadays rather well constrained 
through the global analyses \cite{MRS03,CTEQ6,CTEQ61}, which use the DGLAP \cite{DGLAP} 
$Q^2$-evolution, sum rules and a large amount of data from deep inelastic lepton--proton 
scattering (DIS) and high energy proton--(anti)proton collisions. 
The success of the forthcoming Large Hadron Collider (LHC) program in the search 
for the Higgs boson and physics beyond the Standard Model depends on 
the precision of the PDFs.  

At collider energies, hard processes are abundantly available also in heavy--ion collisions. 
These processes play an important role in testing QCD dynamics and factorization,
as well as in the search of quark-gluon plasma signatures and in the determination of the QCD matter 
properties.  Similar to the free proton case, the computation of nuclear hard process cross sections 
requires the nuclear parton distributions (nPDFs) as input. Thus, there is an obvious need for the 
global analyses of the nPDFs such as presented in 
\cite{Eskola:1998iy,Eskola:1998df,Hirai:2001np,Hirai:2004wq,deFlorian:2003qf}. 

Hard partonic processes taking place at mid-rapidities in nuclear collisions at the Relativistic 
Heavy Ion Collider (RHIC; $A$+$A$ and d+Au at $\sqrt s_{NN}= 200$ GeV) typically probe the nPDFs in a 
kinematic region where the nuclear effects remain relatively small and are fairly well constrained by 
the global analyses. Towards smaller scales and off mid-rapidity, however, the probed region extends 
towards smaller momentum fractions $x$ where both the nuclear effects and the uncertainties in the 
nPDFs grow larger.  Soon at the Large Hadron Collider (LHC; Pb+Pb at $\sqrt s_{NN}= 5.5$ TeV) the 
range of scales and fractional momenta probed will be widened further, both towards smaller $x$ and 
towards larger $Q^2$. This, together with the fact that the nuclear gluon distributions are still 
relatively badly known, emphasizes the importance and topicality of the global analyses in pinning 
down the nPDFs and their uncertainties. 

The fact that nuclear and free proton PDFs are mutually different has 
been known for well over twenty years; for a recent review, see Ref. \cite{Armesto:2006ph}. 
The nuclear effects, the nuclear modifications relative to the free proton PDFs, 
are usually named according to the observed behaviour of the nucleus-to-Deuterium ratio of 
the structure functions $F_2^{A}$ in different $x$-regions,  as follows:
(i)~shadowing; a depletion at $x \lsim 0.1$,
(ii)~antishadowing; an excess at $0.1 \lsim x \lsim 0.3$, 
(iii)~EMC effect; a depletion at $0.3 \lsim x \lsim 0.7$ and 
(iv)~Fermi motion; an excess towards $x\rightarrow1$ and beyond. 
This nomenclature will be used in this paper as well.
The dynamical origin of these nuclear modifications has been actively studied in different frameworks 
as well, see the Refs. e.g. in \cite{Armesto:2006ph,Arneodo:1992wf,Accardi:2004be}.
The DGLAP evolution of the nPDFs and their modifications relative to the free proton PDFs
have been studied for two decades, see e.g. Refs. 
\cite{Qiu:1986wh,Frankfurt:1990xz,Eskola:1992zb,Kumano:1992ef, Kumano:1994pn,Indumathi:1996pb,
Indumathi:1996ky,Frankfurt:2003zd,Accardi:2004be}. 

In a global DGLAP analysis the nPDFs are pinned down as model-independently as possible
at a chosen initial scale on the basis of DGLAP evolution, sum rules and hard process data from 
nuclear collisions.
So far, three groups have presented global DGLAP analyses of the nPDFs analogous to those of the free 
proton.
These are the ones by us, 
Eskola {\it et al.}  {\it EKS98} \cite{Eskola:1998iy,Eskola:1998df}, 
by Hirai {\it et al.}  {\it HKM} \cite{Hirai:2001np} and {\it HKN} \cite{Hirai:2004wq}, 
and by de~Florian and Sassot {\it nDS} \cite{deFlorian:2003qf}.  
The {\it EKS98} analysis \cite{Eskola:1998iy,Eskola:1998df} was the first one to show that a good 
overall fit to the nuclear DIS and Drell-Yan (DY) data can be obtained in a DGLAP-based global 
analysis. In particular, the scale-dependence of the ratio $F_2^{\rm Sn}/F_2^{\rm C}$ observed by the 
NMC experiment \cite{Arneodo:1996ru} was very nicely reproduced by tuning the initial gluon 
modifications suitably. The iterative $\chi^2$ minimization in {\it EKS98} was carried out manually 
(by eye), 
and no well-controlled error estimates were obtained. Since then, extensive further work has been 
done by Kumano and his collaborators in estimating these uncertainties 
\cite{Hirai:2001np,Hirai:2004wq}, and by de~Florian and Sassot \cite{deFlorian:2003qf} in bringing 
the global nPDF analysis to the next-to-leading order (NLO) level.

In this paper, we perform a global analysis of the nPDFs in the {\it EKS98} framework. 
Our study is partly a reanalysis of {\it EKS98} as we take some
guidelines from this old fit. To minimize the number of fit parameters, however, we now apply simpler 
piecewise analytical shapes for the nuclear effects at the initial scale. We also construct the 
nuclear quark modifications in a more transparent way  than in our previous work. The goal 
here is twofold: on one hand, by making the $\chi^2$ minimization procedure automated, we wish to 
check whether the goodness of the old {\it EKS98} fit could still be improved, and on the other hand 
we wish to get a better hold on the uncertainties of the nPDFs, of the gluons in particular, in this 
framework.

The results of this study can be summarized as follows: Within the obtained $\chi^2$ and error 
estimates, we conclude that the old {\it EKS98} parametrization still serves very well. Thus, we do 
not release a new parametrization but recommend to use {\em EKS98}. We also demonstrate how the 
small-$x$ nuclear gluon distributions are, in spite of the good overall fit obtained, still not well 
constrained with the currently available nuclear DIS and DY data elsewhere than perhaps at $x\sim 
0.02-0.04$. 
A comparison with the results from the previous global analyses is also shown, demonstrating the 
nPDFs uncertainties concretely. Finally, a special case beyond the  original {\it EKS98} setup, a 
gluon shadowing clearly stronger than that in $F_2^A/F_2^{\mathrm D}$, is considered
and further developments of the analysis by inclusion of RHIC data are discussed.

This paper is organized as follows. 
In Sec.~\ref{sec:framework}, we define nPDFs according to the {\em EKS98} framework 
and introduce the fitting procedure. 
Section~\ref{sec:results}  contains the results of $\chi^2$ minimization and the 
a detailed comparison with the nuclear DIS and DY data. 
Section~\ref{sec:comparison_others} is devoted for the comparison with previous global analyses. 
In Sec.~\ref{sec:errors} we show the results from the error analysis performed and verify the 
validity of {\em EKS98}. 
In Sec.~\ref{sec:strongershad}, we discuss the possibility of a stronger gluon shadowing supported by 
the RHIC data. 
Conclusions and further discussion are given in Sec.~\ref{sec:discussion}. 

\section{The framework}
\label{sec:framework}

\subsection{Definition of nPDFs}
\label{subsec:nPDFs}

As introduced in {\em EKS98} \cite{Eskola:1998iy}, by a nuclear parton distribution function $f_i^A$ 
we refer to the distribution of a parton type $i$ in a proton\footnote{Note that in {\em HKN} a 
slightly different definition is used, see \cite{Hirai:2001np,Hirai:2004wq}.} bound to a nucleus of a 
mass number $A$. We define and parametrize the nuclear modifications relative to the known free 
proton PDFs $f_i$,   
\begin{equation}
  R_{i}^A(x,Q^2)=\frac{f_i^A(x,Q^2)}{f_i(x,Q^2)}. \label{DefineR_f}
\end{equation}
In the {\em EKS98} framework which we adopt here, the PDFs of the bound neutrons are obtained from 
$f_i^A(x,Q^2)$ by assuming isospin symmetry. Thus, e.g. the total $u$-quark distribution in 
a nucleus of a mass number $A$ and a proton number $Z$ becomes $U_A = Zf_u^A + (A-Z)f_d^A$. 
Correspondingly, the lowest-order QCD parton model expression for the $lA$ DIS structure 
function $F_2$ then becomes 
$F_2^A=\sum_Q e_Q^2 [Q_A+\overline{Q}_A]$, where $Q=U,D,S,...$.

The total amount of fit parameters in the initial ratios $R_i^A$ must be limited for obtaining 
converging well-constrained fits. Unfortunately, the variety of the nuclear data is presently 
not enough to pin down each $R_i^A(x,Q_0^2)$ separately. 
Therefore, following the {\em EKS98} procedure, we can include only three different ratios for each 
nucleus
at an initial scale $Q^2=Q_0^2$ where heavy quarks can be neglected: 
The same average modification $R_V^A=(f_{u_V}^A+f_{d_V}^A)/(f_{u_V}+f_{d_V})$ is applied for all 
valence quarks separately (only at $Q_0^2$ however),  the corresponding sea quark average 
modification $R_S^A$ applied for all sea quarks separately (again at $Q_0^2$ only) and $R_G^A$ for 
gluons. 
While this is the best we can do here, we note that the valence u and d quark nuclear modifications 
may in fact well differ from each other -- for a recent study of how large differences between 
$R_{u_V}^A$ and $R_{d_V}^A$ would explain the NuTeV weak-mixing angle anomaly observed in 
$\nu(\bar\nu)$+Fe DIS, see \cite{Eskola:2006ux}. Also, in the sea quark sector, due to their mutually 
differing absolute distributions, it would be natural to expect that the initial s quark 
modifications are not necessarily identical to those of u and d. Without a multitude of further data 
constraints, however, such details cannot be reliably included in a global analysis.

In the original {\em EKS98} analysis \cite{Eskola:1998iy} we first parametrized the DIS structure 
function ratio 
\begin{equation}
R_{F_2}^A(x,Q^2) \equiv \frac{\frac{1}{A}F_2^A(x,Q^2)}{\frac{1}{2}F_2^{\rm D}(x,Q^2)}     
\label{RF2def}
\end{equation}
at the initial scale $Q_0^2$ and then decomposed this into the valence and sea parts. 
The initial gluon modifications were obtained by adding a double
gaussian distribution on the antishadowing peak of the parametrized $R_{F_2}^A$.
In the current analysis we choose a more straightforward procedure by parametrizing directly the
ratios $R_V^A$, $R_S^A$ and $R_G^A$ at $Q_0^2$. 

The initial scale is here chosen to be $Q_0=1.3$ GeV in order to match the
CTEQ6L1 PDF set \cite{CTEQ61}, which we use to calculate the
absolute nuclear PDFs at $Q_0^2$: 
\begin{equation}
f_{i}^A(x,Q_0^2) = R_{i}^A(x,Q_0^2) f_{i}^{\rm CTEQ6L1}(x,Q_0^2).
\end{equation}  
The lowest order DGLAP scale evolution is calculated using the
routine from the CTEQ collaboration \cite{CTEQ_code}
as it provided fast enough evolution for the minimization purposes.

The key constraints for the nPDFs are given by the nuclear hard process data from  
lepton-nucleus DIS and from the DY dilepton production in proton-nucleus collisions.
We utilize the results from the DIS measurements, 
available in the form of ratios over Deuterium and Carbon,
\begin{equation}
 \frac{\frac{1}{A}d\sigma^{lA}/dQ^2dx}{\frac{1}{2}d\sigma^{l{\mathrm D}}/dQ^2dx}
 \,{\buildrel {\rm LO}\over =}\, R_{F_2}^A(x,Q^2),
 \hspace{1cm}
 \frac{\frac{1}{A}d\sigma^{lA}/dQ^2dx}{\frac{1}{12}d\sigma^{l{\mathrm C}}/dQ^2dx}
 \,{\buildrel {\rm LO}\over =}\, \frac{R_{F_2}^A(x,Q^2)}{R_{F_2}^{\mathrm C}(x,Q^2)}, 
 \label{RF2}
\end{equation}
where the LO connection is implied.
The DY data are available in the form of ratios over Deuterium and Beryllium,
\begin{equation}
   \frac{\frac{1}{A}d\sigma_{DY}^{{\mathrm p}A}/dx_2 dQ^2}
        {\frac{1}{2}d\sigma_{DY}^{\mathrm {pD}}/dx_2 dQ^2}
		\,{\buildrel {\rm LO}\over =}\,
		R_{DY}^A(x_2,Q^2),
 \hspace{1cm}
   \frac{\frac{1}{A}d\sigma_{DY}^{{\mathrm p}A}/dx_1 dQ^2}
        {\frac{1}{9}d\sigma_{DY}^{\mathrm{pBe}}/dx_1 dQ^2}
		\,{\buildrel {\rm LO}\over =}\,
		\frac{R_{DY}^A(x_1,Q^2)}{R_{DY}^{\mathrm Be}(x_1,Q^2)}.
		\label{RDY}
\end{equation}
Above,  $Q^2$ is the invariant mass of the dilepton pair and $Q^2= x_1 x_2\sqrt{s}_{NN}$. 
The data included in this study are shown in Table \ref{Table:Data}. 
The small nuclear effects in Deuterium are neglected.
 
\begin{table}
\begin{center}
{\small
\begin{tabular}{lllll}
 Experiment & Process &  Nuclei & datapoints & Ref.\\
\hline
\hline
 SLAC E-139	& DIS				& He(4)/D        &   18 & \cite{Gomez:1993ri}   \\
 NMC 95, reanalysis 	& DIS	& He/D  	  &   16 & \cite{Amaudruz:1995tq} \\
 \\
 SLAC E-139 & DIS				& Be(9)/D        &   17 & \cite{Gomez:1993ri}   \\
 NMC 96    	& DIS				& Be(9)/C        &   15 & \cite{Arneodo:1996rv} \\
 \\
 SLAC E-139 & DIS				& C(12)/D         &    7 & \cite{Gomez:1993ri}   \\
 NMC 95    	& DIS				& C/D         &   15 & \cite{Arneodo:1995cs} \\
 FNAL-E665	& DIS				& C/D         &    4 & \cite{Adams:1995is}   \\
 NMC 95, reanalysis 	& DIS	& C/D   	  &   16 & \cite{Amaudruz:1995tq} \\
 FNAL-E772 	& DY				& C/D         &    9 & \cite{Alde:1990im}    \\
\\
 SLAC E-139	& DIS				& Al(27)/D        &   17 & \cite{Gomez:1993ri}   \\
 NMC 96    	& DIS 				& Al/C        &   15 & \cite{Arneodo:1996rv} \\
\\
 SLAC E-139	& DIS				& Ca(40)/D        &    7 & \cite{Gomez:1993ri}   \\
 FNAL-E665 	& DIS				& Ca/D        &    4 & \cite{Adams:1995is}   \\
 FNAL-E772 	& DY				& Ca/D        &    9 & \cite{Alde:1990im}    \\
 NMC 95, reanalysis 	& DIS	& Ca/D 		  &   15 & \cite{Amaudruz:1995tq} \\
 NMC 96    	& DIS				& Ca/C        &   15 & \cite{Arneodo:1996rv} \\
\\
 SLAC E-139	& DIS				& Fe(56)/D        &   23 & \cite{Gomez:1993ri}   \\
 FNAL-E772 	& DY				& Fe/D        &    9 & \cite{Alde:1990im}    \\
 NMC 96     & DIS				& Fe/C        &   15 & \cite{Arneodo:1996rv} \\
 FNAL-E866	& DY		 		& Fe/Be       &   28 & \cite{Vasilev:1999fa} \\
\\
 SLAC E-139	& DIS				& Ag(108)/D        &    7 & \cite{Gomez:1993ri}   \\
\\
 NMC 96, $Q^2$ dep.    	& DIS	& Sn(117)/C        &  144 & \cite{Arneodo:1996ru} \\
\\
 FNAL-E772 	& DY				& W(184)/D         &    9 & \cite{Alde:1990im}    \\
 FNAL-E866	& DY		 		& W/Be        &   28 & \cite{Vasilev:1999fa} \\
\\
 SLAC E-139	& DIS				& Au(197)/D        &   18 & \cite{Gomez:1993ri}   \\
\\
 FNAL-E665 	& DIS				& Pb(208)/D        &    4 & \cite{Adams:1995is}   \\
 NMC 96    	& DIS				& Pb/C        &   15 & \cite{Arneodo:1996rv} \\
 FNAL-E665 	& DIS, recalc. 		& Pb/C 		  &    4 & \cite{Adams:1995is}   \\

 \hline		   
 total number of  datapoints     &        &     &  514 &                       \\
\end{tabular}
}
\caption[]{\small The data used in this analysis, grouped according to the nuclei measured. The mass 
numbers are given in parentheses. The number of datapoints refers to those falling into the region 
$Q^2\ge Q_0^2$.}
\label{Table:Data}
\end{center}
\end{table}

As will become clear in the error analysis presented in Sec.~\ref{sec:errors},
the available sets of experimental data do not constrain
the distributions of different parton flavours over the whole range of $x$. 
This will be reflected as some assumptions regarding the shape of the ratios which are
basically the same as in our previous {\em EKS98} work. In particular, motivated
by the requirement of a stable evolution (that the nuclear modifications should not change very 
rapidly from their starting values), a saturation (flattening) of the ratios $R_i^A$ 
at $x\to 0$, and a valence quark -like behavior of the sea and gluon
modifications for $x\to 1$ will be assumed. In the following we explain in detail how
the initial parametrization for $R_V^A(x,Q_0^2), R_S^A(x,Q_0^2)$ and $R_V^A(x,Q_0^2)$
was constructed.

\subsection{Fit functions and parameters}
\label{subsec:parameters}

While the basic idea in the global DGLAP analysis is straightforward, it is a surprisingly nontrivial 
task to develop functional forms for the fit functions for the ratios $R_V^A$, $R_S^A$ and $R_G^A$ 
which can be used in the automated $\chi^2$ minimization process in a transparent way. 
To have a better control over the multidimensional parameter space and 
over the numerical results obtained, each parameter should preferably have a clear interpretation, 
too. Due to the various $A$ and $x$ dependent nuclear effects discussed above and also due to the 
mutual differences between the valence, sea and gluon modifications, the fit functions must contain 
sufficiently many parameters to secure enough flexibility necessary for obtaining good fits. At the 
same time, the number of parameters has to be reduced to a minimum in order to obtain converging fits 
with the rather limited set of data constraints at our disposal. Finally, once the working functional 
forms have been verified, one needs to analyze (on the basis of the data constraints and $\chi^2$ 
fits) which parameters can be left free and which can be fixed.   
Furthermore, the best local minimum in $\chi^2$ has to be verified by optimizing the the initial 
values of all free parameters. All this implies extensive manual labour, even though the actual 
search for the $\chi^2$ minimum is automated.

For the controllability discussed above, and after various other attempts, we ended up constructing 
each of the initial ratios $R_V^A$, $R_S^A$ and $R_G^A$ from three different pieces:
$R_1^A(x)$ at small values of $x$ below the antishadowing\footnote{For antiquarks $R_S^A<1$, by 
antishadowing we refer to the shape similar to $R_{F_2}^A$.} maximum,  $x \le x_a^A$; 
$R_2^A(x)$ in the medium-$x$ region from the antishadowing maximum to the EMC minimum, 
$x_a^A \le x \le x_e^A$; and 
$R_3^A(x)$ in the Fermi-motion region in the large-$x$ region,
$x\ge x_e^A$;

\begin{eqnarray}
  &&R_1^A(x) = c_0^A+(c_1^A+c_2^A x)[\exp(-x/x_s^A) 
           - \exp(-x_a^A/x_s^A)],   \qquad x\le x_a^A \label{R1}\\
  &&R_2^A(x) = a_0^A + a_1^A x + a_2^A x^2 + a_3^A x^3, 
             \qquad x_a^A \le x \le x_e^A\\ \label{R2} 
  &&R_3^A(x) = \frac{b_0^A-b_1^A x}{(1-x)^{\beta^A}},      
             \qquad x_e^A \le x \label{R3}.
\end{eqnarray}

In choosing the above forms, we were motivated by the functional forms used before in  
Hard Probes \cite{HPC} (see \cite{Eskola:2002us}), {\em EKS98} \cite{Eskola:1998iy} and {\em HKN} 
\cite{Hirai:2004wq}. 
Matching is done by requiring continuity of the fit functions and setting their first derivatives to 
zero at $x_a^A$ (local maximum) and $x_e^A$ (local minimum). 
As the coefficients $a_i^A$, $b_i^A$ and 
$c_i^A$ are somewhat unintuitive, we shall quote the results in terms of the following more 
transparent set of seven parameters from which these coefficients can be easily solved:

\begin{tabular}{ll}
  $y_0^A$          & $R_1^A$ at $x\rightarrow 0$,\\
  $x_s^A$          & a slope factor in the exponential, \\
  $x_a^A$, $y_a^A$ & position and height of the antishadowing maximum \\
  $x_e^A$, $y_e^A$ & position and height of the EMC mimimum \\
  $\beta^A$		   & slope of the divergence of $R_3$ at $x\rightarrow 1$.
\end{tabular} \\

Each of the above parameters is in principle yet specific to a nucleus $A$. 
This (at least) doubles the amount of parameters. We parametrize the 
$A$-dependence in a simple power-like form:
\begin{equation}
  z_i^A = z_i^{A_{\rm ref}} (\frac{A}{A_{\rm ref}})^{\,p_{z_i}},
  \label{eq:Adependence}
\end{equation}
where $z_i = x_s, x_a, y_a\ldots$, and choose the reference nucleus 
to be Carbon, $A_{\rm ref}=12$. The number of parameters
we have for the valence, sea and gluon ratios each is thus 14: 
the Carbon parameters (suppressing the superscript C to lighten the notation)
$y_0$, $x_s$, $x_a$, $x_e$, $y_a$, $y_e$, $\beta$, and their 
powers $p_{y_0}$, $p_{x_s}$, $p_{x_a}$, $p_{x_e}$, $p_{y_a}$, $p_{y_e}$ and $p_{\beta}$.
Altogether this makes $3\times14=42$ free parameters. Even if the momentum and 
baryon number conservation, imposed individually for each nucleus, reduce this 
number by four, it is clearly far too large
for a converging $\chi^2$ minimization process, given the limited data constraints 
we have. In order to radically reduce the number of free parameters, we proceed as follows, keeping 
in mind the focus on the small- and medium-$x$ regions.

\begin{itemize}

\item{Fermi-motion.}
In the large-$x$ region, where valence quarks dominate, 
the DIS or DY data do not give proper constraints for gluons or sea quarks. 
Thus, we fix the Fermi-motion slopes $\beta^A$ in $R_S^A$ and $R_G^A$ to be the same as in $R_V^A$. 
Based on our previous  {\em EKS98} work, we fix $\beta=0.3$ and $p_{\beta}=0$ in $R_V^A$, thus   
ignoring a possible $A$-dependence of $\beta^A$.

\item{EMC effect.}
Gluons originate from valence quarks at small scales and large $x$. Therefore, they should reflect 
the EMC effect observed in $R_V^A$ ($R_{F_2}^A$). From the gluons the effect should then be 
transmitted on to $R_S^A$ as well. We have checked that this is indeed the case in the DGLAP 
evolution \cite{Heli_thesis}. Thus, by assuming the similarity of the EMC-minima in each initial 
ratio $R_G^A$, $R_S^A$ and $R_V^A$, one reaches a stable scale evolution of this nuclear effect. As 
the available data, however,  constrain the EMC effect in detail only in $R_V^A$, we fix the location 
parameters $x_e^A$ and the magnitude parameters $y_a^A$ of the EMC-minima in $R_G^A$ and $R_S^A$ to 
be identical to those in $R_V^A$. For the valence part, we noticed that allowing for an $A$ 
dependence in $x_e^A$ did not improve the overall fits, hence we fix $p_{x_e}=0$ for simplicity.

\item{Antishadowing.}
In course of the present analysis we also noticed that the location parameters $x_a^A$ of the 
antishadowing maxima in $R_V^A$ and in $R_S^A$ typically become almost $A$-independent and that the 
weak $A$ dependence does not improve the obtained fits. We therefore set $p_{x_a}=0$ in $R_V^A$ and 
$R_S^A$. In order to reduce the number of gluon parameters to the very minimum, we simply fix 
$x_a^A$ of gluons to be identical to that in valence but leave $y_a$ and $p_{y_A}$ free for 
controlling the height of the antishadowing maximum in an $A$-dependent way.

\item{Shadowing.}
In the small-$x$ parts, based on $\chi^2$-checks, we drop the $A$-dependence of the slope parameters 
$x_s^A$, hence setting $p_{x_s}$ to zero and and leaving $x_s$ free in all ratios. 

\item{Conservation laws.} 
Baryon number and momentum conservation are used to calculate $y_0^A$ for $R_V^A$ 
and $R_G^A$, respectively, for each nucleus individually. This eliminates the parameters $y_0$ and 
$p_{y_0}$ for the valence and gluon modifications. For the sea quarks, these parameters are left 
free.
\end{itemize}

\paragraph{}All this brings the number of free parameters down to 16: 
$x_s$, $x_a$, $x_e$, $y_a$, $p_{y_a}$, $y_e$ and $p_{y_e}$ in $R_V^A$; 
$y_0$, $p_{y_0}$, $x_s$, $x_a$, $y_a$ and $p_{y_a}$ in $R_S^A$; and 
$x_s$, $y_a$, and $p_{y_a}$ in $R_G^A$. 
Table \ref{Table:Params} summarizes the above discussion on the parameters as well as their values 
obtained in finding a "best" local minimum with respect to the fit parameters for 
\begin{equation}
\chi^2 = \sum_{i=1}^{N_{\mathrm{data}}} \left( \frac{\mathrm{data}_i-\mathrm 
{theory}_i}{\Delta_i}\right )^2.
\end{equation}
As the data errors $\Delta_i$, we take the given statistical and systematic errors added in 
quadrature.

Some remarks on the functional form adopted for the shadowings at small-$x$ are in order here.  Since 
the valence modification $R_V^A$ is rather well constrained by the DIS and DY data in the large- and 
medium-$x$ regions, its small-$x$ behaviour becomes relatively stringently constrained by the baryon 
number sum rule. Unfortunately, in the absence of DIS (or DY) data for $R_{F_2}^A$ at $x<0.001$ in 
the DGLAP region $Q^2\gsim 1$~GeV$^2$, the sea quark $R_S^A$ and the gluon $R_G^A$ cannot be pinned 
down similarly well in the small-$x$ region -- thus their behaviour and error estimates at small $x$ 
are bound to be specific to the fit function forms assumed.

The motivation for choosing the smallest-$x$ form of $R_1^A(x)$ in Eq.~(\ref{R1}), where shadowing 
levels off to a constant value at $x=0$, is the fact that such saturation of shadowing has been 
observed in the very small-$x$ \& very small-$Q^2$ DIS data (see Fig.~10 in \cite{Arneodo:1995cs}) 
and the fact that the $Q^2$ dependence there is rather weak (see Figs.~11 and 12 in 
\cite{Arneodo:1995cs}). In  doing this, however, we should keep in mind that the implications of the 
observed saturation of shadowing are not clear for the nPDFs at perturbative scales: power 
corrections $\sim (Q^2)^{-n}$ \cite{Qiu:2003vd} are most likely important in the DIS cross sections 
at small enough scales, and also nonlinearities \cite{GLRMQ} (neglected here) are expected to play a 
role in the scale evolution at sufficiently small-$x$ \& small-$Q^2$.

In the previous {\em EKS98} analysis, due to the modest and non-negative $\log Q^2$-slopes of 
$R_{F_2}^A$ discussed above, we fixed the smallest-$x$ behaviour of $R_{F_2}^A(x,Q_0^2)$ to a value 
slightly above the saturation of shadowing observed at lower scales.  
The $\log Q^2$ slopes of $R_{F_2}^A$ computed from the DGLAP equations at small $x$ 
\cite{Prytz:1993vr,Eskola:1998iy,Eskola:2002us}, 
\begin{eqnarray}
 \frac{\partial R_{F_2}^A(x,Q^2)}{\partial \log Q^2}
 &\propto&
 \alpha_s\frac{xg(2x,Q^2)}{F_2^D(x,Q^2)}
 \biggl\{R_G^A(2x,Q^2)-R_{F_2}^A(x,Q^2)\biggr\},
 \label{Saturation}
\end{eqnarray}
are non-negative if $R_G^A(2x)\ge R_{F_2}^A(x)$. 
In {\em EKS98}, it was shown that an ansatz 
$R_G^A(x\rightarrow 0)\rightarrow R_{F_2}^A(x\rightarrow 0)$ works well for the smallest $x$.
In the present analysis, we want to test the above {\em EKS98} gluon framework and thus keep the 
saturation of gluon shadowing independent of that in $R_S^A$.

\section{Results}
\label{sec:results}

\subsection{Final parameters and their interpretation}
\label{subsec:khi2}
In minimizing the $\chi^2$ with respect to the free parameters, 
we used the MINUIT routines from the CERN Program Library  \cite{MINUIT}. 
Only after reducing the number of free parameters down to 16, 
and after extensive searches for suitable initial parameter values, 
we were able to find a converging fit indicating a local minimum 
of the $\chi^2$. The obtained parameters for the best fit found are shown in Table
\ref{Table:Params}. The resulting goodness of the fit was $\chi^2 = 410.15$ for $N=514$
data points and 16 free parameters, giving 
$\chi^2/N=0.80$ and $\chi^2/\mathrm{d.o.f.} = 0.82$. 

\begin{table}[tbh]
\begin{center}
\begin{tabular}{l|c|lll}
   & Param.   	&  Valence      	&  Sea           	&  Gluon \\
\hline
\hline
 1 &  $y_0$    	& {baryon sum}	&  0.88909       	&  {momentum sum}	\\
 2 &  $p_{y_0}$ & {baryon sum}	&   -8.03454$\times 10^{-2}$  	& {momentum sum}  	\\ 
 3 &  $x_s$    	&  0.025 ($l$) 		&  0.100 ($u$)	&  0.100 ($u$)   \\
 4 &  $p_{x_s}$ &  0, {fixed}   &  0, {fixed}   &  0, {fixed}   \\
 5 &  $x_a$    	&  0.12190      	&  0.14011       	&  {as valence} \\
 6 &  $p_{x_a}$ &  0, {fixed}   &  0, {fixed}   &  0, {fixed}   \\
 7 &  $x_e$    	&  0.68716      	&  {as valence} &  {as valence} \\
 8 &  $p_{x_e}$ &  0, {fixed}   &  0, {fixed}   &  0, {fixed}   \\
 9 &  $y_a$    	&  1.03887      	&  0.97970       	&  1.071 ($l$)    \\
 10&  $p_{y_a}$ &  1.28120$\times 10^{-2}$   	& -1.28486$\times 10^{-2}$   	&  3.150$\times 10^{-2}$ ($u$)  		\\
 11&  $y_e$    	&  0.91050      	&  {as valence} &  {as valence} \\
 12&  $p_{y_e}$ & -2.82553$\times 10^{-2}$   	&  {as valence} &  {as valence} \\
 13&  $\beta$  	&  0.3          	& {as valence}  &  {as valence} \\
 14& $p_{\beta}$&  0, {fixed}   & {as valence}  &  {as valence}	\\
\hline
\end{tabular}                                                  \\
($u$) upper limit; ($l$) lower limit

\caption[]{\small List of all parameters defining the modifications $R_V^A$, $R_S^A$ and $R_G^A$ in 
Eqs.~(\ref{R1}-\ref{R3}) at the initial scale $Q_0^2=1.69$~GeV$^2$. The parameters 
$y_0$, $y_a$, $y_e$, $x_s$, $x_a$, $x_e$ and $\beta$ are for the reference nucleus $A=12$, 
and the powers $p_i$ define the $A$-dependence in the form of Eq.~(\ref{eq:Adependence}).
The obtained final results for the fitted 16 free parameters are shown and the fixed parameters are 
indicated.  The parameters which drifted to their upper (u) and lower (l) limits are indicated, see the text 
for details. 
}
\label{Table:Params}
\end{center}
\end{table}

As indicated in the table, the parameters $x_s$ controlling the slopes of $R_1^A$ near the 
antishadowing region were drifting to their limits. In spite of various attempts we failed to 
improve upon this unwanted feature. Obviously, there is still room for developing the chosen 
functional forms in the quark sector too.
However, as the fits obtained now (and already in {\em EKS98}) are very good, new functional forms 
are not likely to improve the $\chi^2$ essentially. In fact, this was our observation also at 
different stages of the present analysis: in spite of the non-converging fits often obtained (which 
were due to too many free parameters allowed or badly guessed initial parameter values), the obtained 
fits themselves were equally good. 

The gluon sector, however, is the most troublesome one, as all the data constraints are indirect and 
not very conclusive when put into the context of a global analysis: rather large changes in the gluon 
shadowing and antishadowing can be compensated for by fairly moderate modifications in the quark 
sector. As a result, gluons have a minor effect in the overall $\chi^2$.
The gluonic parameters $y_a$ and $p_{y_a}$, which are drifting to their limits (see the table), 
reflect these problems. 

As described above, the functional form $R_1^A$ at very small $x$ preassumes the saturation of 
shadowing also for gluons. 
The height of the antishadowing bump $y_a$ and its $A$-dependence are correlated with the 
parameters $y_0$ and its $A$ dependence $p_{y_a}$ which are computed from the momentum sum rule:
the larger the $y_a$, the smaller the $y_0$. Even though no essential improvement over the 
$\chi^2$ was noticed in varying the limits of $y_a$ and $p_{y_a}$, a clear trend was observed: 
as indicated by reaching the lower limit of $y_a$, the amounts of gluon antishadowing and shadowing 
always tend to be minimized. This in turn means that gluon shadowing saturates at a value larger than 
that of sea quarks and that the $\log Q^2$ slopes of $R_{F_2}^A$ at the smallest $x$ remain positive. 
These observations coincide with the results from previous global analyses {\em HKN} 
\cite{Hirai:2004wq} and {\em nDS} \cite{deFlorian:2003qf}. 

We thus conclude that the present DIS and DY data and the sum rule constraints suggest that gluon 
shadowing is weaker or at most as strong as that in sea quarks. As one of the goals here is to test 
the {\em EKS98} framework for our final results summarized in Table~\ref{Table:Params} we have set 
the lower limits of the free gluonic antishadowing parameters $y_a$ and $p_{y_a}$ in such a way that 
the gluon shadowing levels off to the same value as that of sea quarks ($R_{F_2}^A$). 
The benefit in doing this is that we can keep the {\em EKS98}-like good agreement with the clearly 
positive $\log Q^2$ slopes of $F_2^{\rm Sn}/F_2^{\rm C}$ observed at $x\sim 0.01$, see 
Fig.~\ref{Fig:RF2SnC} ahead.

As explained above, in the present analysis the valence and gluon parameters $y_0^A$ are computed 
from baryon number and momentum sum rules, correspondingly, for each nucleus separately. For 
completeness, we note that a power-law fit of Eq.~(\ref{eq:Adependence}) to the values obtained, 
using $A=12$ and 208, gives $y_0=0.9288$ and $p_{y_0}=-0.031209$
for valence and $y_0=0.8898$ and $p_{y_0}=-0.084315$ for gluons. With such parametrization, baryon 
number and momentum would be conserved with sufficient accuracy, within a few per cent, for all 
nuclei.

\begin{figure}[hbt]
\centering\includegraphics[width=16cm]{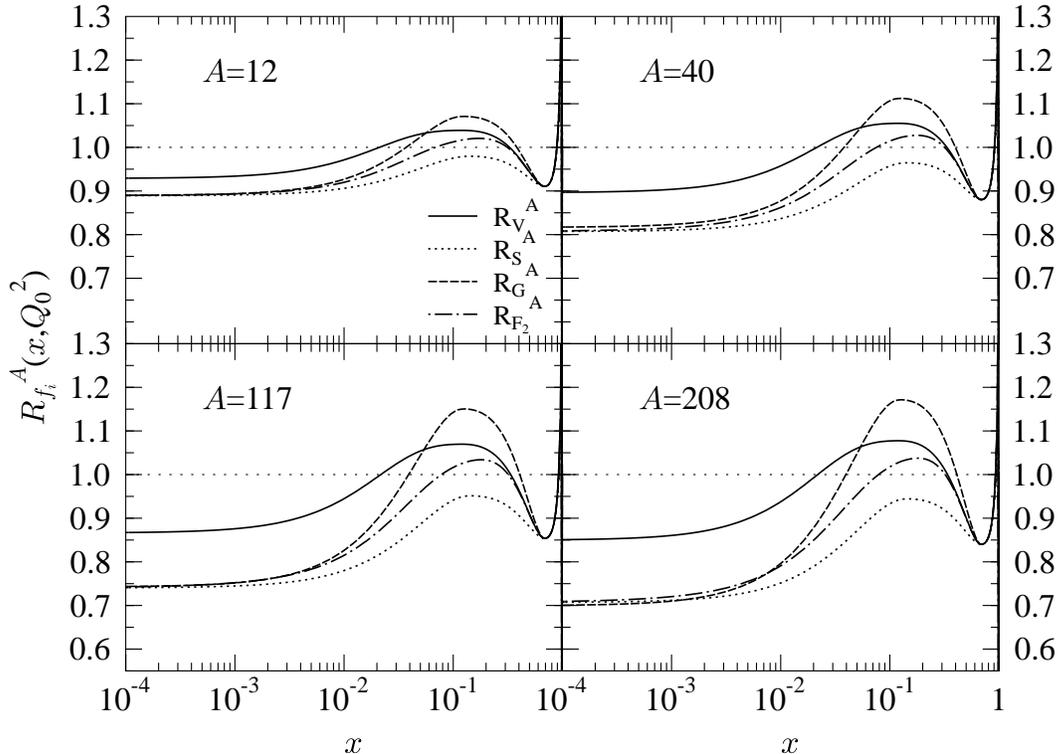}
\caption[]{\small Initial nuclear ratios $R_V^A(x,Q_0^2)$ (solid lines),
  $R_S^A(x,Q_0^2)$ (dotted lines), $R_G^A(x,Q_0^2)$ (dashed lines) and
  $R_{F_2}^A(x,Q_0^2)$ (dotted-dashed lines) for $A=12$, 40, 117 and 208
  at $Q_0^2=1.69$ GeV$^2$.}
\label{Fig:Initial}
\end{figure}

The obtained initial nuclear modifications are shown in Fig.~\ref{Fig:Initial}, where
we plot $R_V^A(x,Q_0^2)$ (solid lines), $R_S^A(x,Q_0^2)$ (dotted
lines), $R_G^A(x,Q_0^2)$ (dashed lines) and $R_{F_2}^A(x,Q_0^2)$
(dotted-dashed lines) for nuclei $A=12$, 40, 117 and 208 at an initial
scale $Q_0^2=1.69$ GeV$^2$.

\begin{figure}[!]
\centering\includegraphics[width=15cm]{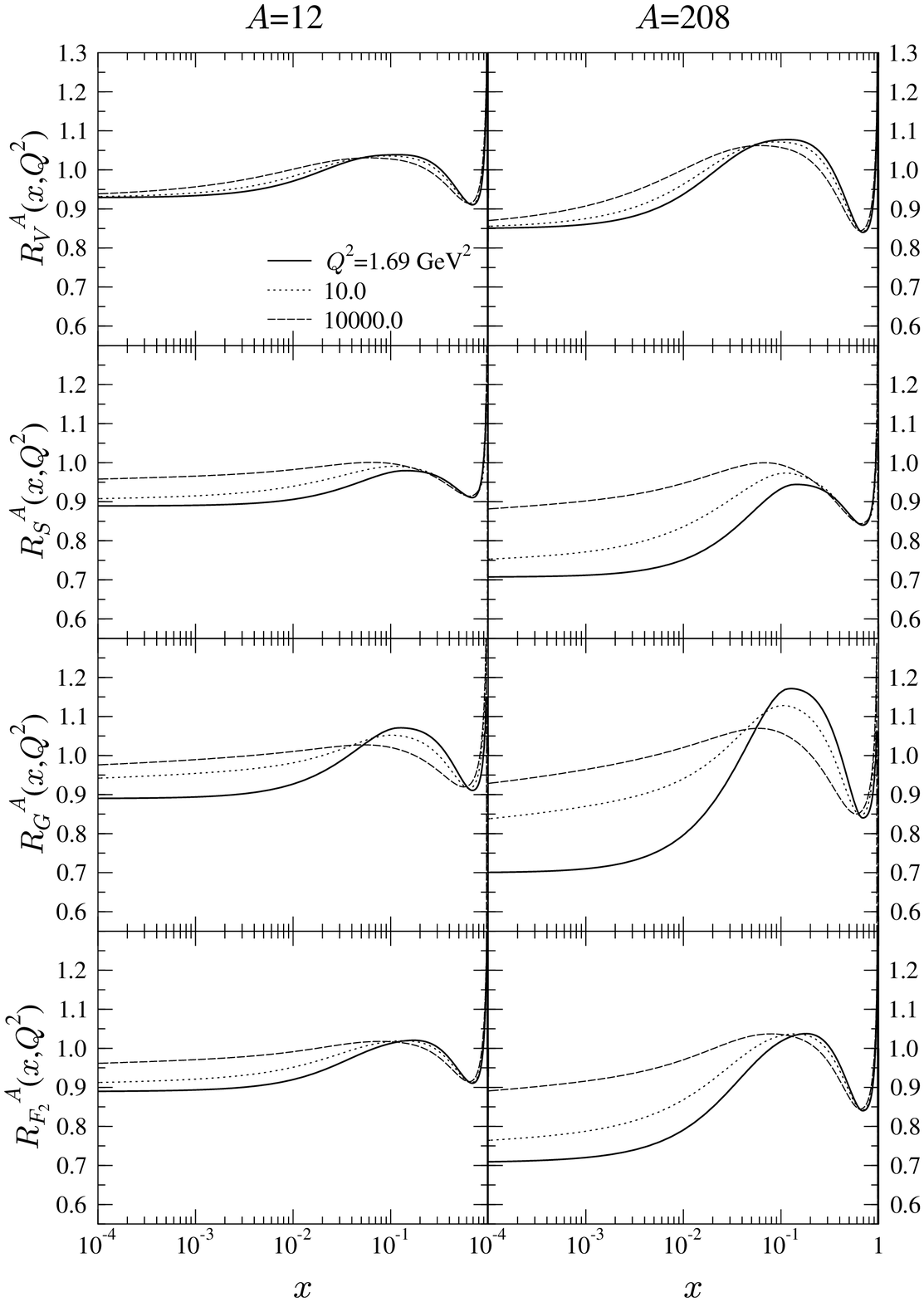}
\caption[]{\small Scale evolution of nuclear modifications: the ratios 
  $R_V^A(x,Q^2)$, $R_V^A(x,Q^2)$, $R_V^A(x,Q^2)$, and $R_{F_2}^A(x,Q^2)$ at scales 
  $Q^2=1.69$, 100 and 10000 GeV$^2$ for $A=12$ and 208.}
\label{Fig:RAQ2}
\end{figure}

The scale evolution of the nuclear effects is shown in Fig.
\ref{Fig:RAQ2}, where the ratios are plotted for $A=12$ and $A=208$ as a function of $x$,
at fixed scales $Q^2=Q_0^2=1.69$ GeV$^2$ (solid), 10 GeV$^2$
(dotted) and $10^4$ GeV$^2$ (dashed). In the regions where no stringent data constraints 
are available for sea quarks and gluons, notice the systematic scale dependence
at small $x$ ($\log Q^2$ slopes do not change their sign), and the stability of the ratios
near the EMC minimum.
 
\subsection{Comparison with data}
\label{subsec:data}

\begin{figure}[!]
\vspace{-.0cm}
\centering\includegraphics[width=15cm]{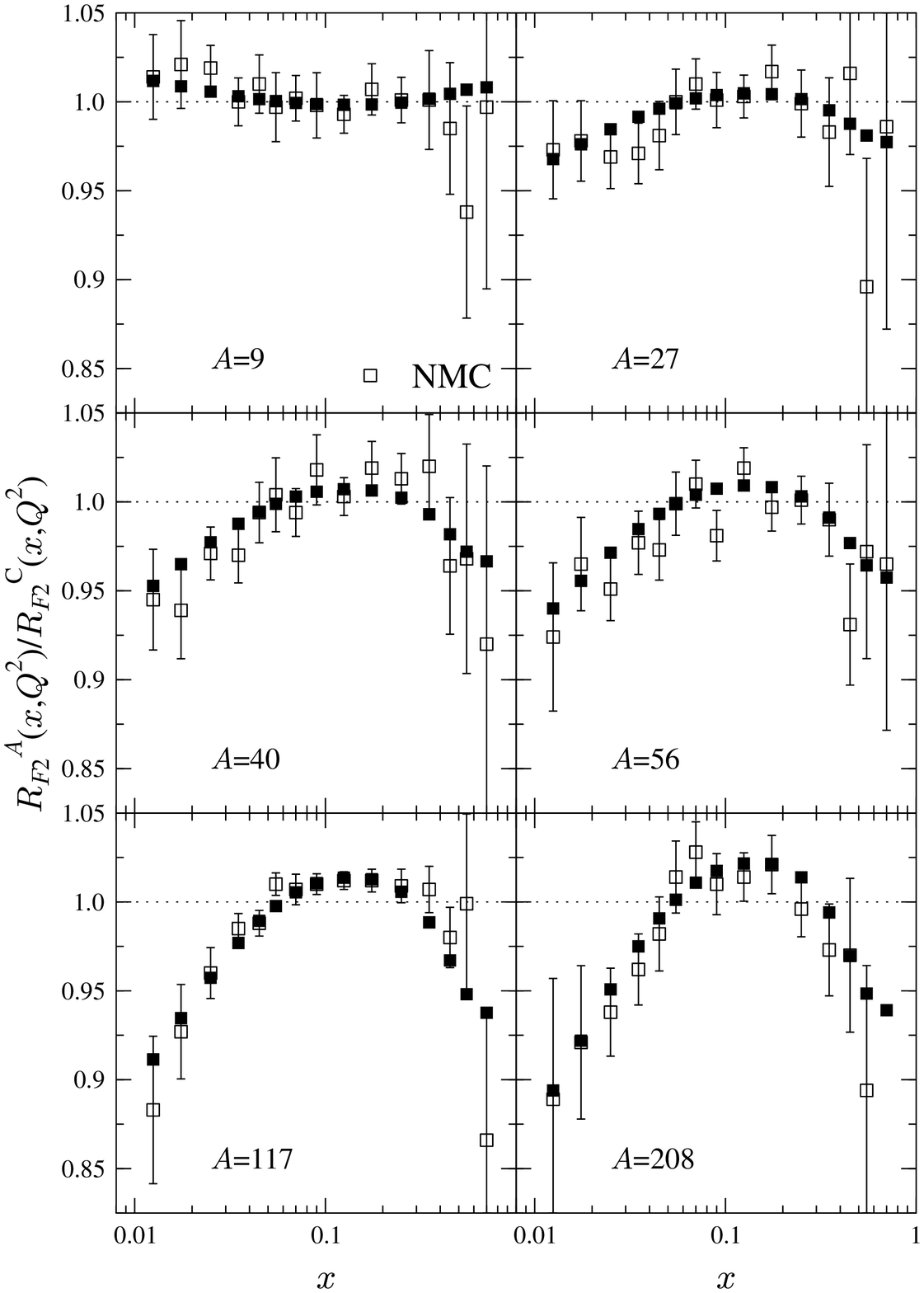}
\caption[]{\small The computed ratio $R_{F_2}^A(x,Q^2)$ vs. $R_{F_2}^{\mathrm
    C}(x,Q^2)$ compared with the NMC data \cite{Arneodo:1996rv}. The open
  symbols are the data points with errors added in quadrature, the filled
  ones are the corresponding results from this analysis.}
\label{Fig:RF2AC}
\end{figure}

\begin{figure}[!]
\vspace{-1cm}
\centering\includegraphics[width=12cm]{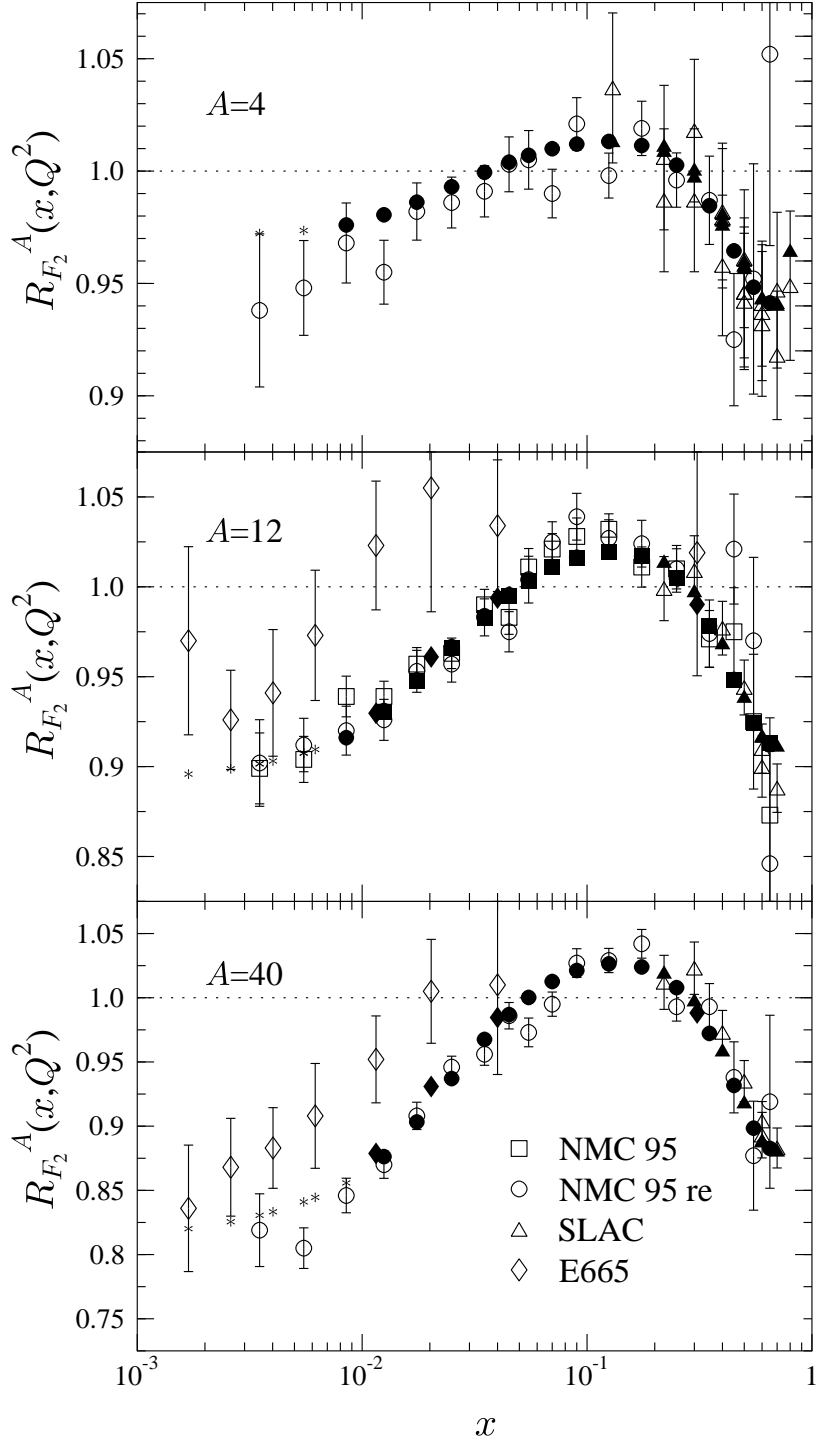}
\caption[]{\small Calculated $R_{F_2}^A(x,Q^2)$ (filled symbols) are
  compared to SLAC (triangles) \cite{Gomez:1993ri}, E665 (diamonds)
  \cite{Adams:1995is}, NMC 95 (squares) \cite{Arneodo:1995cs} 
  and reanalysed NMC 95 (circles) data
  \cite{Amaudruz:1995tq}.  The asterisks denote our results
  calculated at the initial scale $Q_0^2$, these are for the smallest-$x$ data points 
  whose scales lie in the region $Q^2<Q_0^2$.}
\label{Fig:RF2A1}
\end{figure}

\begin{figure}[!]
\begin{center}
\includegraphics[width=12cm]{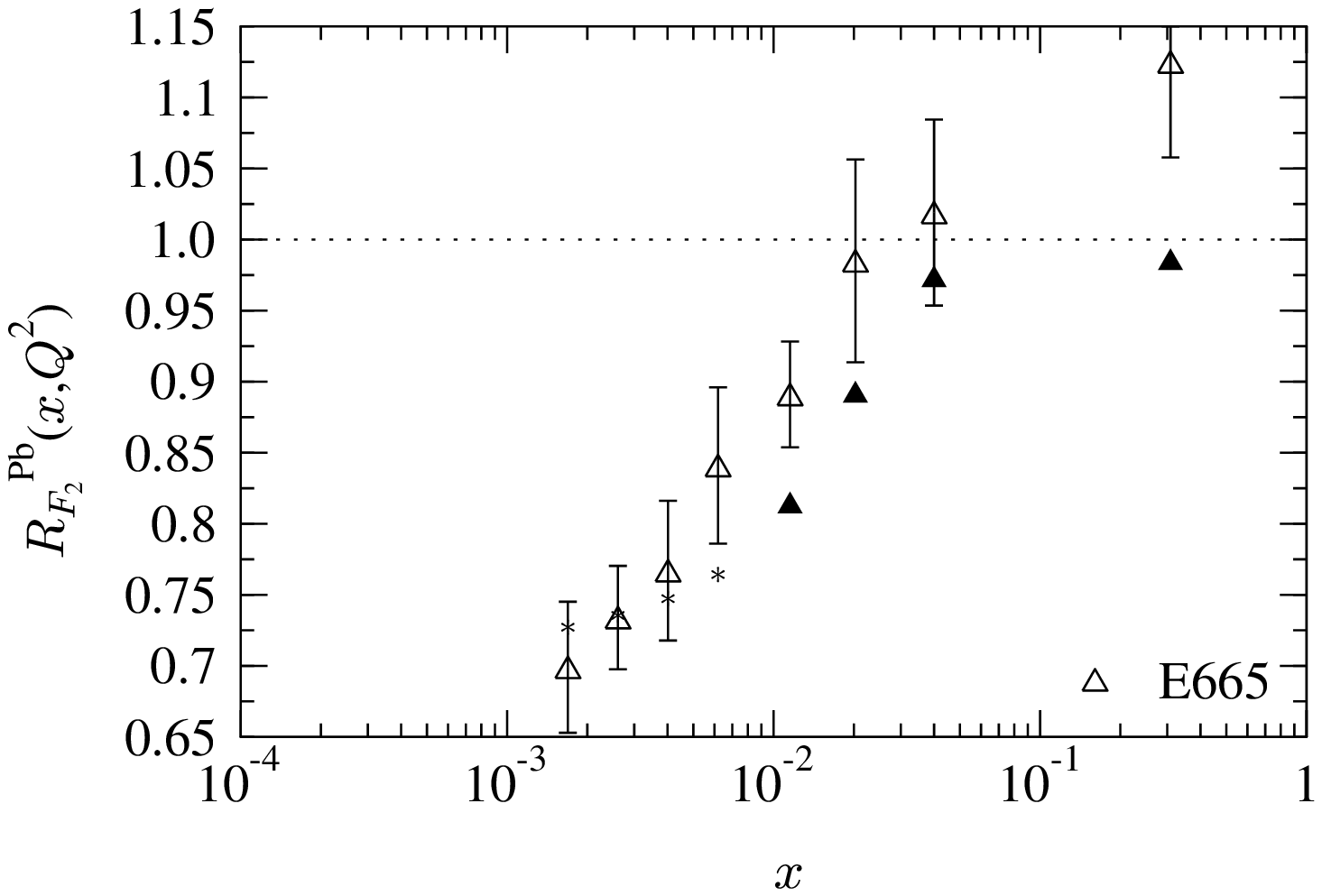}
\vspace{0.5cm}
\includegraphics[width=12cm]{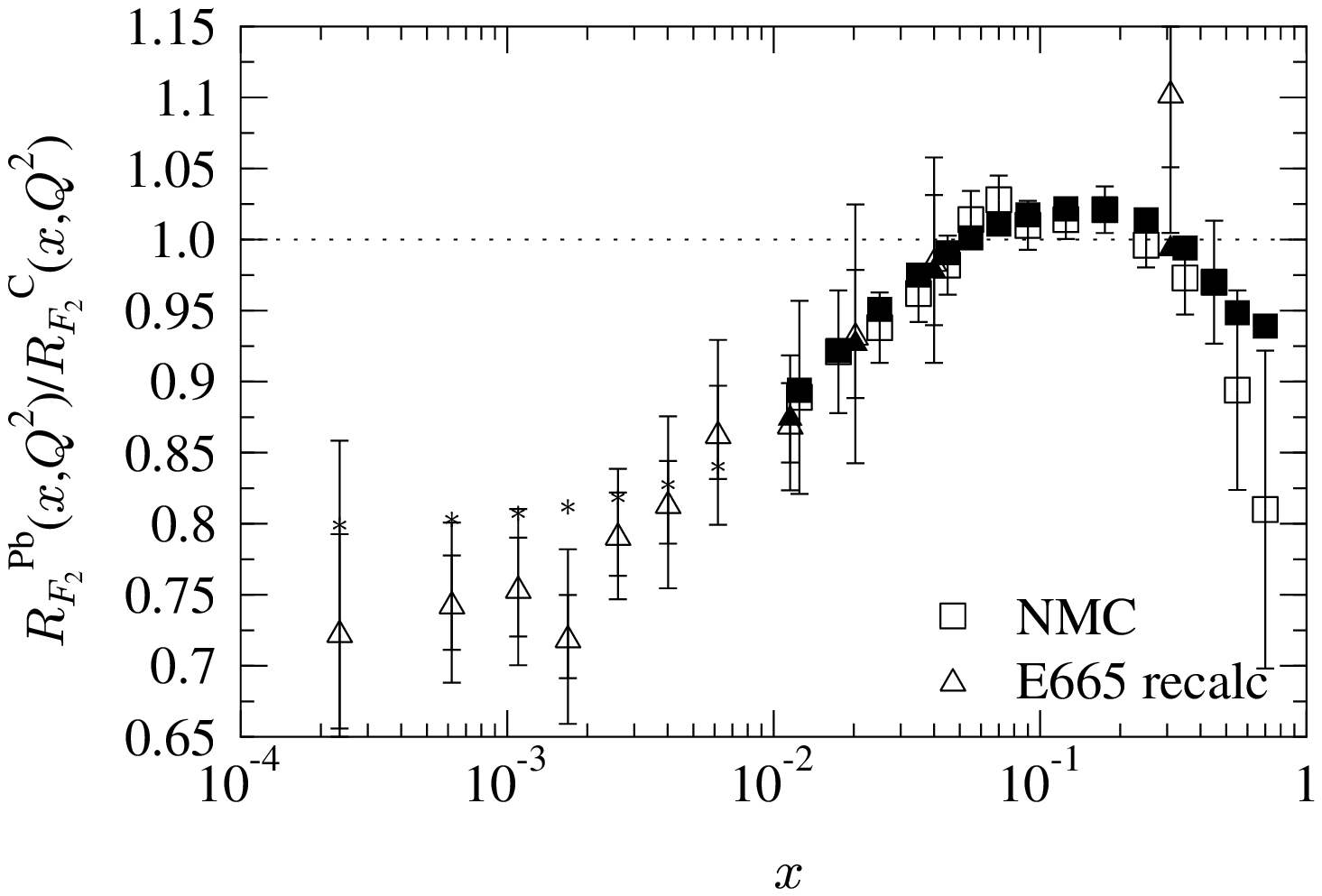}
\caption[]{\small 
  {\bf Top:} The ratios $R_{F_2}^{\mathrm{Pb}}/R_{F_2}^{\mathrm D}$ 
  from the E665 experiment (open triangles) \cite{Adams:1995is} 
  compared with the results from the present analysis (filled triangles).  
  {\bf Bottom:}
  Comparison of the ratios $R_{F_2}^{\mathrm{Pb}}/R_{F_2}^{\mathrm C}$. 
  The NMC data \cite{Arneodo:1996rv} are shown by open squares, the 
  ratios calculated from the E665 data \cite{Adams:1995is} by open triangles.
  For the error estimates in the latter case, see the text. The corresponding 
  theoretical results are again shown by the filled symbols, and by asterisks if the 
  experimental $Q^2$ is below our initial scale $Q^2_0$. 
   }
\label{Fig:RF2PbCD}
\end{center}
\end{figure}

\begin{figure}[!]
\centering\includegraphics[width=14cm]{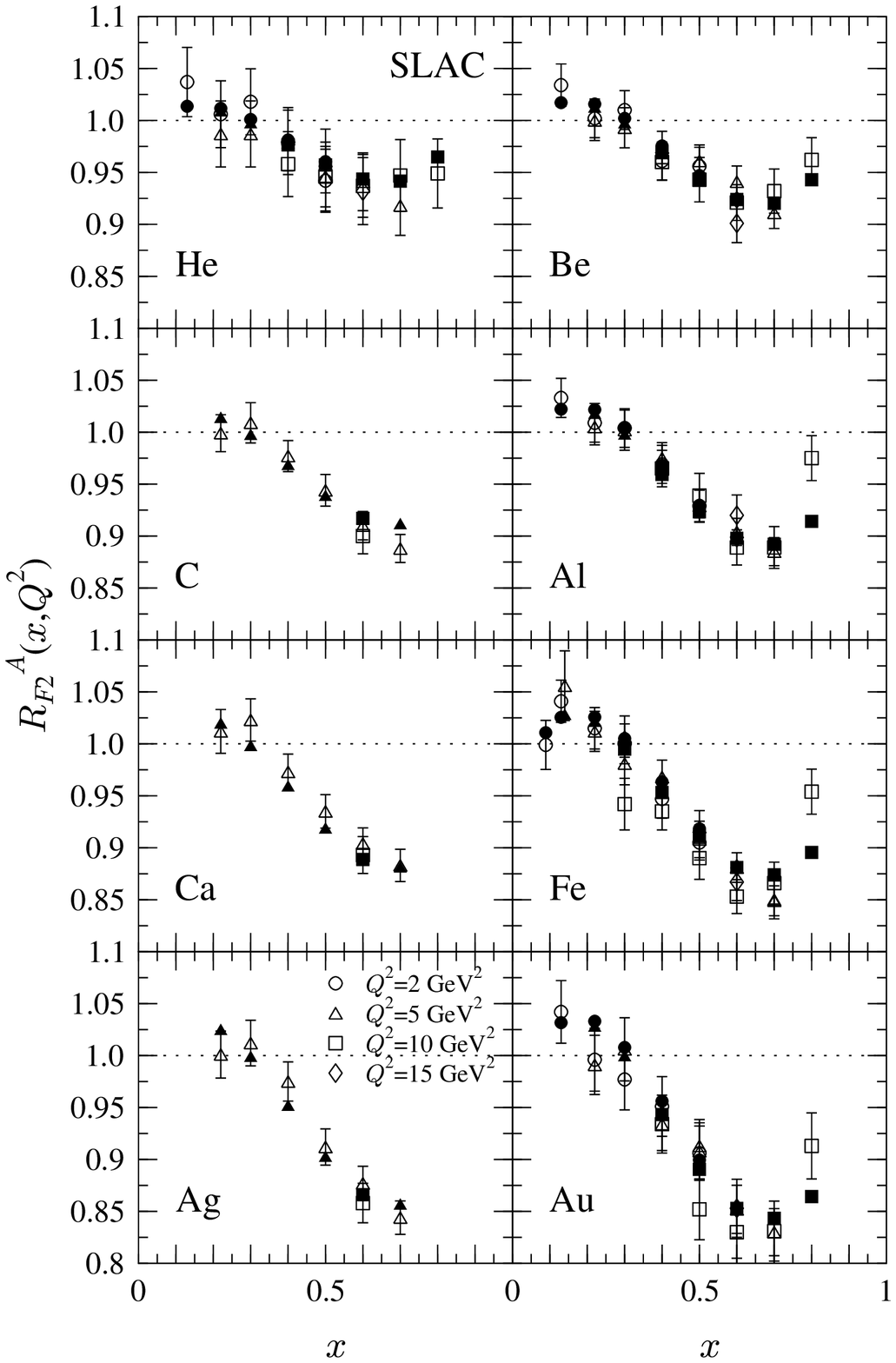}
\caption[]{\small  The calculated ratio $R_{F_2}^A(x,Q^2)$ compared with the  
  SLAC data \cite{Gomez:1993ri}. Data points at $Q^2=2$ GeV$^2$ are shown
  by circles,  $Q^2=5$ GeV$^2$ by triangles,  $Q^2=10$
  GeV$^2$ by squares and $Q^2=15$ GeV$^2$ by diamonds. The corresponding 
  filled symbols mark our results.}
\label{Fig:RF2SLAC}
\end{figure}

\begin{figure}[htb]
\centering\includegraphics[width=13cm]{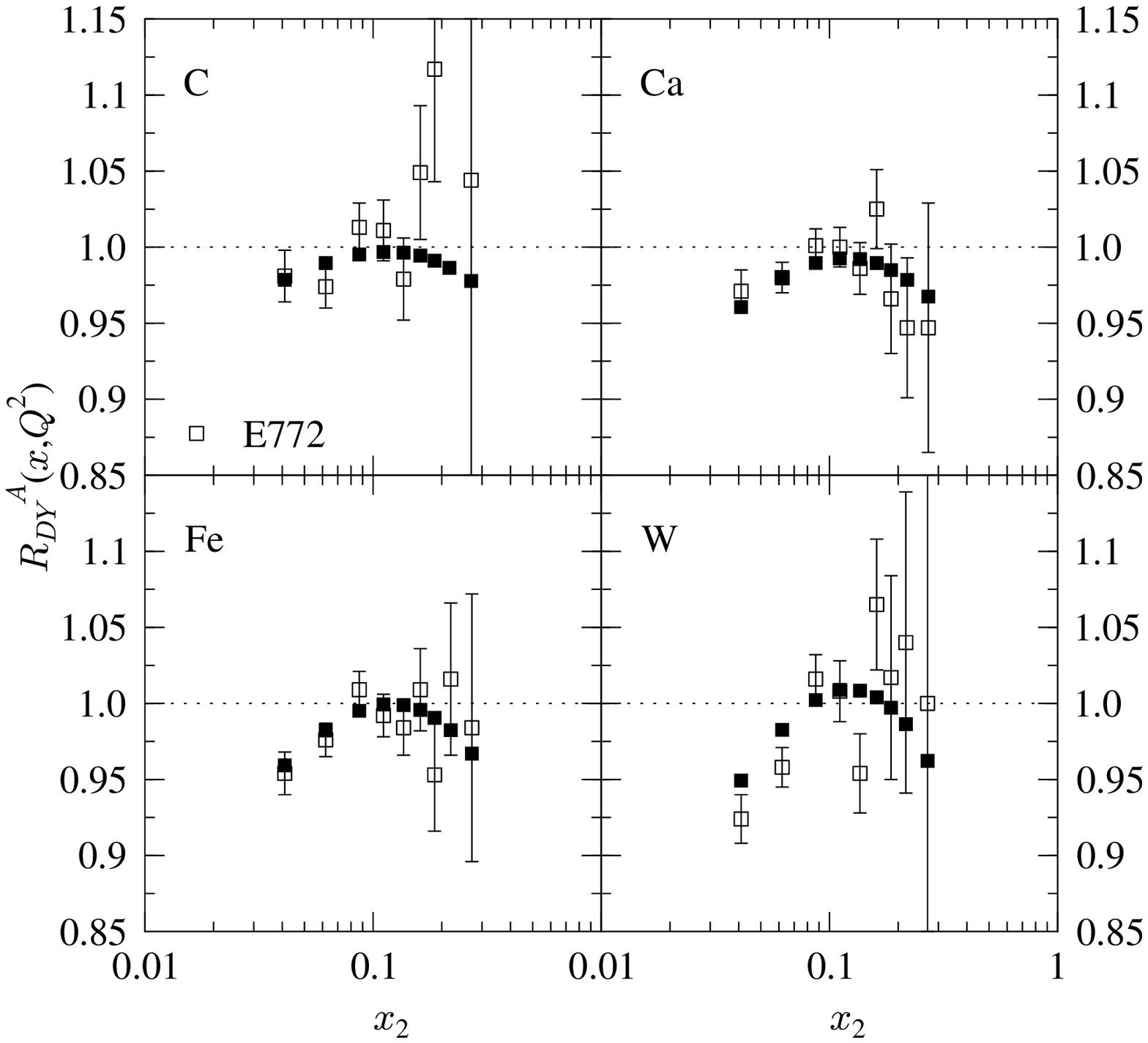}
\caption[]{\small The ratio of the computed LO differential Drell-Yan cross sections (open squares),
  $(d\sigma^{{\mathrm p}A}/dQ^2dx_2)/(d\sigma^{\mathrm{pD}}/dQ^2dx_2)$, and 
  the E772 data \cite{Alde:1990im} (filled squares).}
\label{Fig:E772}
\end{figure}

\begin{figure}[!]
\centering\includegraphics[width=14cm]{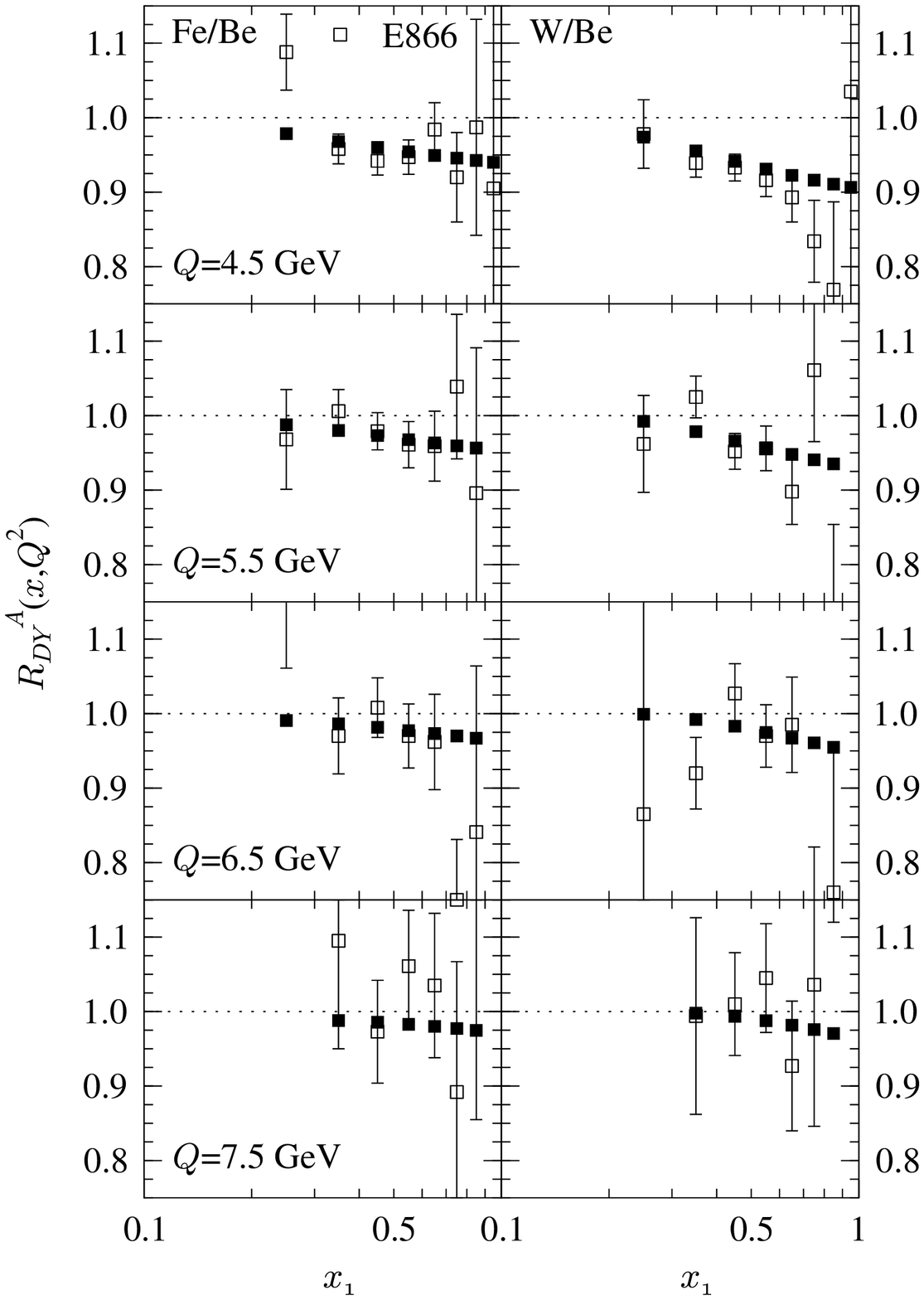}
\caption[]{\small The ratio of the computed LO differential Drell-Yan cross sections (open squares),
  $(d\sigma^{{\mathrm p}A}/dQ^2dx_1)/(d\sigma^{\mathrm{pD}}/dQ^2dx_1)$, compared with the 
  E866 data  \cite{Vasilev:1999fa} as a function of $x_1$ at four different invariant mass ($Q^2$)
  bins. Some data points lie outside the shown region;
  nevertheless their error bars are shown if they extend to the figure.
}
\label{Fig:E886}
\end{figure}

\begin{figure}[!]
\centering\includegraphics[width=15cm]{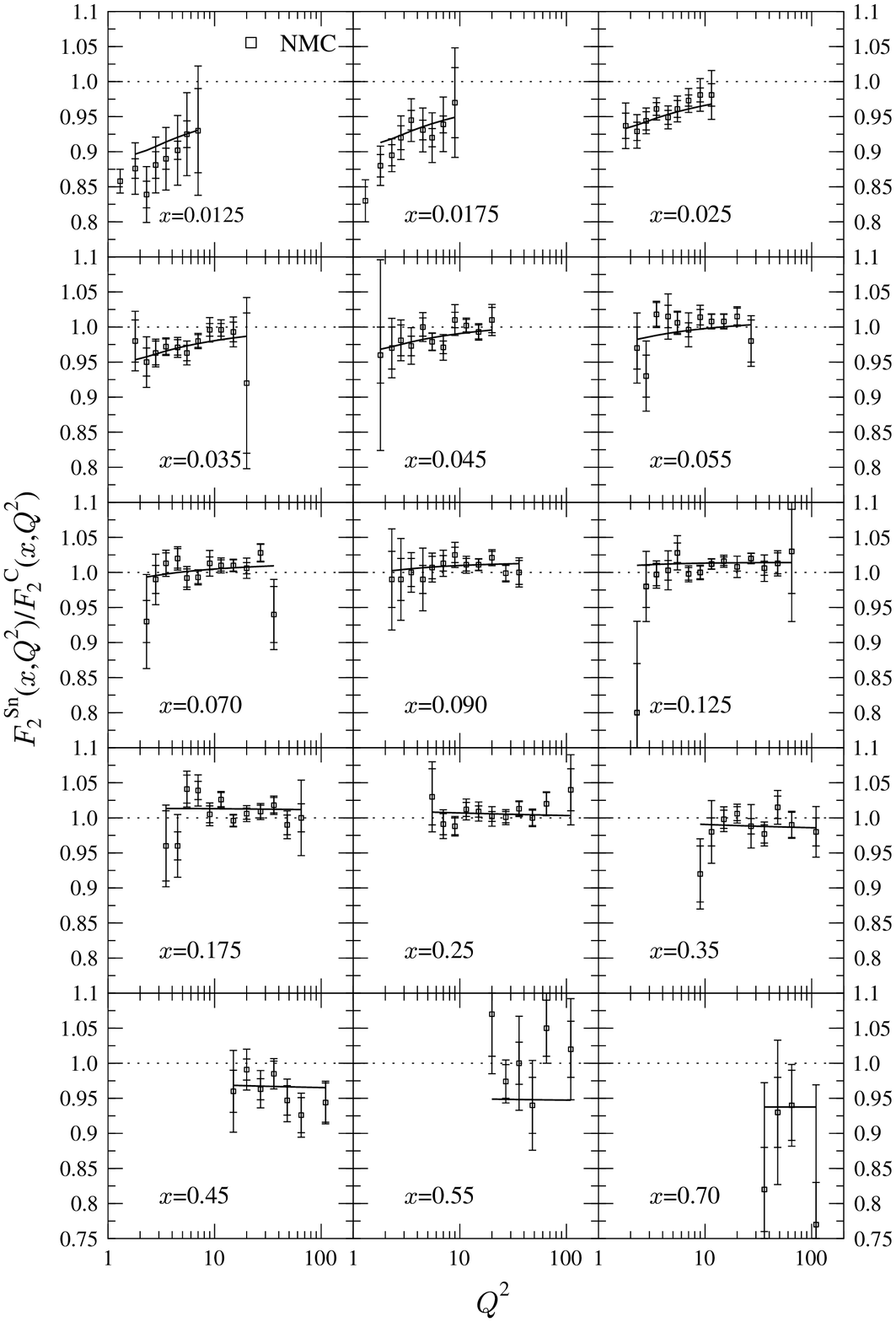}
\vspace{-0.5cm}
\caption[]{\small The calculated scale evolution (solid lines) of the ratio
  $F_2^{\mathrm{Sn}}/F_2^{\mathrm{C}}$ compared with the NMC data
  \cite{Arneodo:1996ru} for  several fixed values of $x$. The inner error bars are the statistical 
ones, the outer ones stand for the statistical and systematic errors added in quadrature.}
\label{Fig:RF2SnC}
\end{figure}

Next we compare the obtained results with the data included in the analysis 
and illustrate the good overall agreement obtained. 
The DIS data can be found in Figs.~\ref{Fig:RF2AC}-\ref{Fig:RF2SLAC}
and in \ref{Fig:RF2SnC}, and the DY data in Figs.~\ref{Fig:E772}-\ref{Fig:E886}.
In the plots below, the statistical and systematic errors of the data have been 
added in quadrature.

In Fig. \ref{Fig:RF2AC} we show the computed ratio 
$\frac{1}{A}F_2^A/\frac{1}{12}F_2^{\mathrm C} = R_{F_2}^A/R_{F_2}^{\mathrm C}$ 
against the  NMC data \cite{Arneodo:1996rv} for various nuclei. 
The open squares are the NMC data points 
and the filled squares are our results computed at the corresponding values of $x$ and
$Q^2$. This data set plays a major role in constraining the $A$-systematics of nuclear quark 
distributions at small $x$.

In Fig.~\ref{Fig:RF2A1} we compare the computed ratio $R_{F_2}^A(x,Q^2)$ with 
the data from SLAC \cite{Gomez:1993ri}, E665 \cite{Adams:1995is}, NMC 95 \cite{Arneodo:1995cs} and NMC 95 \cite{Amaudruz:1995tq} reanalysis. The open triangles, diamonds, squares and circles stand for the data and 
the corresponding filled symbols show our results. Note that at the same/similar values of $x$ the 
values of $Q^2$ can vary between the different data sets, hence the multiple filled symbols at these 
$x$. In the figure, we have also included the small-$x$ 
data points whose $Q^2$-values lie below our initial scale. The asterisks show our results
at our $Q^2_0$. To compare these points with the data, one should perform the scale evolution
downwards. We do not consider this here (and hence these data points are not included in the 
$\chi^2$ minimization either) but from the figure we can immediately see, as 
the $\log Q^2$ slopes of $R_{F_2}^A$ are positive and modest, and as the points computed at a higher 
scale lie above the NMC data, that the agreement is good 
also in that part of the small-$Q^2$ region where the DGLAP might still be valid.

Similar comparisons are shown in Fig. \ref{Fig:RF2PbCD} for the ratios
$R_{F_2}^{\mathrm{Pb}}/R_{F_2}^{\mathrm D}$ and
$R_{F_2}^{\mathrm{Pb}}/R_{F_2}^{\mathrm C}$.  
In the upper panel we show the ratio $R_{F_2}^{\mathrm{Pb}}/R_{F_2}^{\mathrm D}$
from the E665 experiment (open triangles) \cite{Adams:1995is}. 
The agreement is not very good, which is not surprising as the NMC and E665 
data sets in Fig.~\ref{Fig:RF2A1} do not agree, either (the NMC data has more weight in the analysis 
due to their smaller error bars). However, as noticed by 
the NMC well in the past \cite{Arneodo:1996rv}, if one considers the ratio of ratios, 
$R_{F_2}^{\mathrm{Pb}}/R_{F_2}^{\mathrm C}$,
the agreement between these data sets becomes very good. This is shown in the lower panel of Fig.~
\ref{Fig:RF2PbCD}, where we plot the data from NMC (open squares) \cite{Arneodo:1996rv}
and together with a ratio calculated  from the 
E665 (open triangles) data for $R_{F_2}^{\rm Pb}$ and $R_{F_2}^{\rm C}$  \cite{Adams:1995is}.
We obtain the error bars for the computed E665 Pb/C ratio by first adding the statistical and
systematic errors in quadrature separately for Pb/D and C/D, and
then taking these errors to be independent. The filled squares and triangles again show our 
DGLAP results corresponding to the data points, while the asterisks mark our results at the 
$x$-points where our initial scale is higher than the $Q^2$ in the E665 data. 

Further comparison with the SLAC data \cite{Gomez:1993ri} for
$R_{F_2}^A(x,Q^2)$ are shown in Fig.  \ref{Fig:RF2SLAC} for various
nuclei and $Q^2$ scales. This set of data plays an important role in constraining
$x$- and $A$-dependence of the valence quark distributions in the EMC region. 
The filled symbols again stand for our results, the open ones for the data.

Figure \ref{Fig:E772} shows the comparison of the calculated LO Drell-Yan
cross section ratios, Eq.~(\ref{RDY}), to the FNAL E772 data \cite{Alde:1990im}. 
The momentum fraction $x_2$ is that of the nuclear parton.
Open squares with error bars present the data points and filled squares
show the calculated values. As can be seen, the calculated values fit
the data rather well, except at the smallest $x_2$-points for tungsten 
(for which the {\em EKS98} seems to work slightly better).

Figure \ref{Fig:E886} then shows the comparison with a newer
E866 data set \cite{Vasilev:1999fa} on the DY ratio 
$(d\sigma^{{\mathrm p}A}/dQ^2dx_1)/(d\sigma^{\mathrm{pD}}/dQ^2dx_1)$ 
as a function of the projectile-parton momentum fraction.
Four different invariant mass bins are considered.
Large values of $x_1$ now correspond to small values of $x_2$. Confirming the trend seen in the 
previous figure,  we note that the $A$ dependence of shadowing could be slightly stronger in order to 
better match with the DY data. Within the present global analysis, however, we were unable to improve 
on this feature. 

Finally, in Fig. \ref{Fig:RF2SnC} we plot the scale evolution of the
ratio $\frac{1}{117}F_2^{\mathrm{Sn}}/\frac{1}{12}F_2^{\mathrm C}$ compared with 
the data from NMC \cite{Arneodo:1996ru} for several fixed values of $x$. 
The $\log Q^2$ slopes of the data at small $x$, which are sensitive to 
the gluon modifications as shown in Eq. (\ref{Saturation}) are reproduced very well, 
similar to {\em EKS98}. Note that the 15 panels here correspond to the 15 data points 
in the lower left panel of Fig.~\ref{Fig:RF2AC}, so that the normalization of 
$\frac{1}{117}F_2^{\mathrm{Sn}}/\frac{1}{12}F_2^{\mathrm C}$ at each $x$ is given by the overall fit.
Thus in the upper left panel ($x=0.0125$) of Fig.~\ref{Fig:RF2SnC} the normalization is slightly 
higher than that of the data, while in the third panel ($x=0.025$) both the normalization and the 
$\log Q^2$ slopes match perfectly.

The NMC data at the smallest-$x$ panels of Fig.~\ref{Fig:RF2SnC} play an important role in 
constraining the nuclear gluon modifications. These data were the key ingredient in the {\em EKS98} 
analysis in pinning down the nuclear gluon modifications around $x\sim 0.03$, for more discussion see 
also
\cite{Eskola:2002us}. We note, however, that in an automated global analysis like we perform here, 
this role becomes not quite as clear: even relatively large variations of the gluon modifications 
induce changes practically only in the first few panels of this figure. The weight that these these 
panels have in the $\chi^2$ is rather small among the $~500$ other data points from cross sections 
mostly sensitive to the changes in the quark sector. 

\section{Comparison with previous analyses}
\label{sec:comparison_others}

Table~\ref{Table:Khi2} summarizes the $\chi^2$ obtained in this work, {\em EKS98}  
\cite{Eskola:1998iy,Eskola:1998df}, {\em HKM}  \cite{Hirai:2001np}, {\em HKN}  \cite{Hirai:2004wq} 
and {\em nDS} \cite{deFlorian:2003qf} analyses. Since each analysis uses different initial scales, 
different amount of data points and different data sets, we quote the values given in the original 
references (except for {\em EKS98} whose $\chi^2$ we compute here using CTEQ6L1). As seen in the 
table, the goodness of the fit using the {\em EKS98} nuclear effects is very close to the one 
obtained in this work and also (contrary to the claim in \cite{deFlorian:2003qf}) quite close to the 
good fit obtained in the LO analysis {\em nDS}. Interestingly, the $\chi^2$ of the NLO fit of {\em 
nDS} is slightly smaller than the LO ones, lending further support to the validity of the global 
analysis.

\begin{table}[tbh]
\begin{center}
\begin{tabular}{llcccccc}
 Set  &Ref.		&$Q_0^2/$GeV$^2$	& $N_{\mathrm{data}}$	& $N_{\mathrm{params}}$	&  $\chi^2$	& 
$\chi^2/N$& $\chi^2/$d.o.f.\\
\hline
\hline
 This work 	 &							 &1.69 	& 514	& 16	& 410.15	& 0.798		& 0.824\\
 {\em EKS98} &\cite{Eskola:1998iy}		 &2.25	& 479	& --	& 387.39	& 0.809		& --  \\ 
 {\em HKM}   &\cite{Hirai:2001np}		 &1.0	& 309	& 9		& 546.6		& 1.769		& 1.822\\
 {\em HKN}   &\cite{Hirai:2004wq}		 &1.0	& 951	& 9     &1489.8		& 1.567		& 1.582\\
 {\em nDS}, LO  &\cite{deFlorian:2003qf} & 0.4  & 420	& 27	& 316.35	& 0.753		& 0.806\\
 {\em nDS}, NLO &\cite{deFlorian:2003qf} & 0.4  & 420	& 27	& 300.15	& 0.715		& 0.764\\
 \hline
\end{tabular}                                                  \\

\caption[]{\small The goodness of the fits obtained in different global analyses. 
}
\label{Table:Khi2}
\end{center}
\end{table}

\begin{figure}[tbh]
\centering\includegraphics[width=15cm]{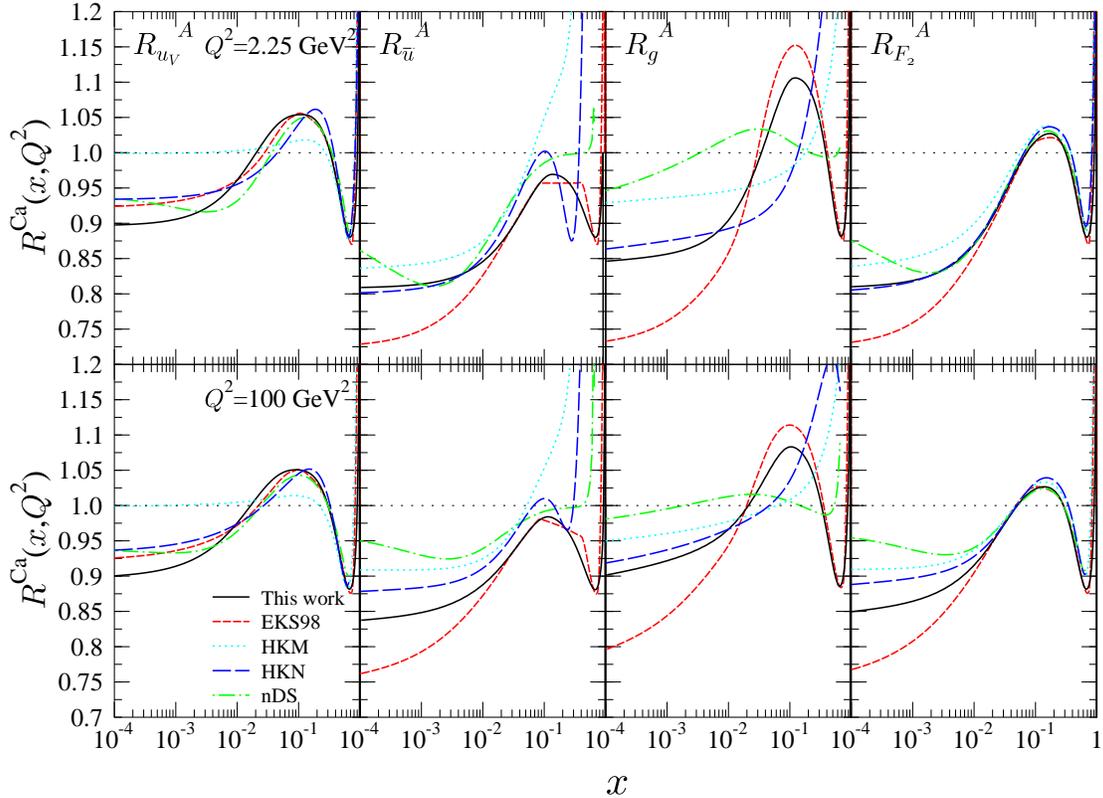}
\caption[]{\small (Colour online) Comparison of different nPDF modifications for Ca:
  This work, {\em EKS98} \cite{Eskola:1998iy,Eskola:1998df}, {\em HKM}
  \cite{Hirai:2001np}, {\em HKN} \cite{Hirai:2004wq} and {\em nDS} (NLO)
  \cite{deFlorian:2003qf} are plotted at scales $Q^2=2.25$ and 100
  GeV$^2$. Calcium is used here as it is fairly well constrained by 
  the data.}
\label{Fig:CompRfA}
\end{figure}

To demonstrate the remaining uncertainties in the nPDFs,  
we show in Fig.  \ref{Fig:CompRfA} the comparison between this work (solid),
{\em EKS98}  \cite{Eskola:1998iy,Eskola:1998df} (dashed), 
{\em HKM}  \cite{Hirai:2001np} (dotted), {\em HKN}  \cite{Hirai:2004wq} (long-dashed) and 
{\em nDS} (NLO) \cite{deFlorian:2003qf} (dot-dashed) sets. 
The ratios $R_V^A$, $R_{\bar u}^A$, $R_G^A$ and $R_{F_2}^A$ are plotted for
$A=40$ at scales $Q^2=2.25$ and 100 GeV$^2$.  
We choose Calcium here as there are both small-$x$ and larger-$x$ DIS data 
and DY data available for this nucleus. The lower one of the scales considered 
is the initial scale in the {\em EKS98} set. 

As can be seen in Fig.~\ref{Fig:CompRfA}, the quantitative main difference between the present 
analysis and {\em EKS98} lies in the small-$x$ behaviour of sea quark and gluon modifications. For 
the sea quarks, the difference is merely due to the different form of the fit functions chosen: in 
the present work, shadowing in $R_S^A$, and thus also that in $R_{F_2}^A$, levels off faster. Like in 
the original {\em EKS98} framework, the very-small-$x$ behaviour of $R_G^A$ at $Q_0^2$ is tied to 
that of $R_{F_2}^A$ (but indirectly, through restricting the limits of the free parameters 
controlling the antishadowing maximum), thus also the gluon shadowing saturates now faster than in 
{\em EKS98}, and hence we have also somewhat less antishadowing in gluons. Recall also the small 
difference in the initial scales here and in {\em EKS98}. In the region $x\sim 0.02-0.03$, where the 
ratios $R_G^A$ are indirectly constrained by the NMC data in Fig.~\ref{Fig:RF2SnC}, the results from 
the present work and {\em EKS98} are very similar. 

Regarding all sets, we first notice that in the mid/large-$x$ region $x\gsim 0.1$ the ratios 
$R_{F_2}^A$ are almost identical, thanks to the constraints given by the DIS data for the $x$, $Q^2$ 
and $A$ dependence of $R_{F_2}^A$. Since in the large-$x$ region, $x\gsim 0.3$ or so, valence quarks 
dominate $R_{F_2}^A$, also the ratios $R_V^A$ from different sets agree nicely there. 

The role of the DY data in pinning down both $R_V^A$ and $R_S^A$ in the small/mid-$x$ region 
$0.01\lsim x \lsim 0.3$ can be concretely seen in the figure. In the {\em HKM} \cite{Hirai:2001np} 
analysis (dotted lines), the DY data  was not included. As a result, the {\em HKM} fit suggested 
$R_S^A\gg 1$ at $x>0.1$, which in turn compensated the smallness of $R_V^A(x\sim 0.1)$ (see the left 
panel) in reproducing $R_{F_2}^A$. The main improvement from {\em HKM} to {\em HKN} 
\cite{Hirai:2004wq} was the inclusion of the DY data in the fit. This translates into better 
constrains and a better agreement with {\em EKS98} for $R_V$ over the whole $x$-region and also for 
$R_S^A$ at $0.01\lsim x\lsim 0.1$.  The fact that the ratios $R_V$ from different global analyses 
agree so nicely is 
quite reassuring, as it demonstrates that the average valence quark modifications can be pinned down 
in a manner which does not depend much on the specific form chosen for the fit functions.

At $x\gsim 0.2$, where valence quarks start to dominate the quark sector, sea quarks are not 
sufficiently constrained by either DIS or DY data -- hence the large variations in $R_S^A$ from 
set to set. This is the case also in the very-small-$x$ region $x\lsim 0.01$, in the absence of 
sufficient data constraints there. Thus, the very-small-$x$ behaviour of $R_S^A$ is specific to the 
form of the fit function chosen. 

\begin{figure}[tbh]
\centering\includegraphics[width=15cm]{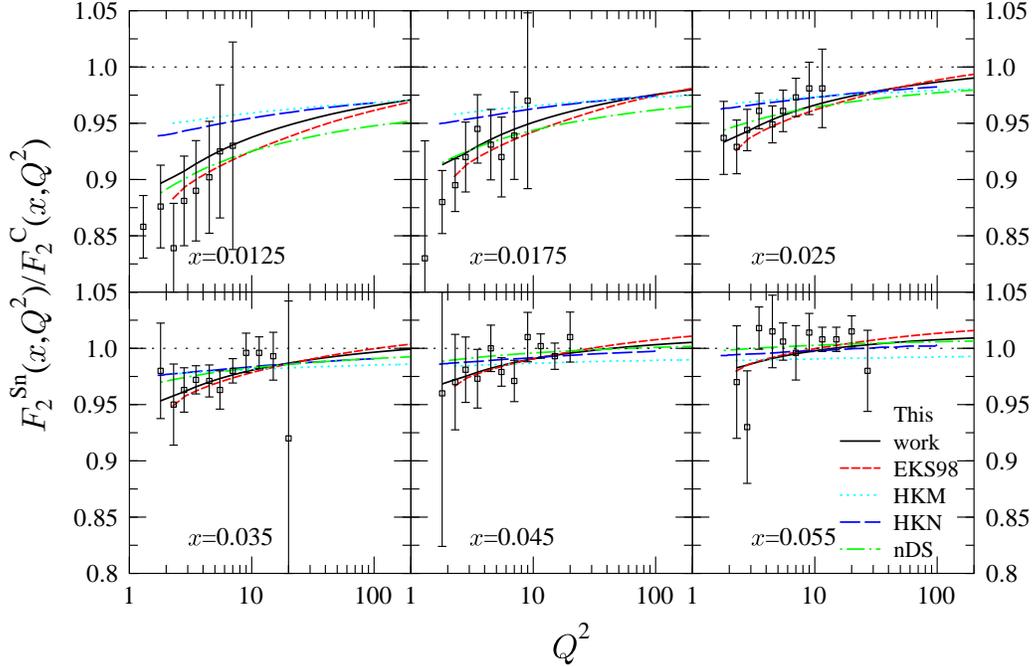}
\vspace{-0.5cm}
\caption[]{\small (Colour online) Comparison of the results from this analysis (solid), 
{\em EKS98} (dashed), {\em HKM} (dotted),  {\em HKN} (long-dashed) and {\em nDS} (dot-dashed) for 
the ratio $\frac{1}{117}F_2^{\mathrm{Sn}}/\frac{1}{12}F_2^{\mathrm{C}}$. As in Fig.~\ref{Fig:RF2SnC}
(6 first panels there), the data is from NMC \cite{Arneodo:1996ru}. 
}
\label{Fig:CompRF2SnC}
\end{figure}

As can be seen in Fig.~\ref{Fig:CompRfA}, the nuclear gluon distributions in general are still quite 
badly constrained, resulting in large differences between the different sets. In the absence of data 
which would sufficiently stringently constrain the gluon modifications over a wide enough $x$-range, 
the results from the global fits are bound to depend on the form of the fit functions chosen. 
To demonstrate this, we replot the ratio 
$\frac{1}{117}F_2^{\mathrm{Sn}}/\frac{1}{12}F_2^{\mathrm{C}}$
in Fig.~\ref{Fig:CompRF2SnC} for the six smallest-$x$ panels of Fig.~\ref{Fig:RF2SnC}. 
As can be seen here, the $\log Q^2$ slopes of $F_2^{\mathrm{Sn}}/F_2^{\mathrm{C}}$ become 
flatter in {\em HKN} and {\em HKM} than those in the present analysis, 
{\em EKS98} and {\em nDS}. The reason for this can be seen from the ratios 
$R_G^A$ at $x\gsim 0.02-0.04$ in Fig.~\ref{Fig:CompRfA} and from Eq.~(\ref{Saturation}): 
the larger $R_G^A$ is relative to $R_{F_2}^A$, the faster is the $Q^2$ dependence of $R_{F_2}^A$.
However, as commented in the previous section, the small-$x$ NMC data which would give at least some 
constraints for the gluons at $x\sim 0.02-0.04$, has a relatively small weight in the global 
analysis. All this makes it difficult to pin down the nuclear gluon modifications.

\section{Error Analysis}
\label{sec:errors}

Next, to quantify the above discussion on the uncertainties, we proceed to  
the error analysis, one of the goals in the present paper.
We do this by using the Hessian method, which is
one of the standard methods in multiparameter analyses as it takes the
parameter correlation into account. The error matrix, or the Hessian
matrix, is the inverse of the second derivative matrix of the fitting
function $\chi^2$ with respect to its free parameters. The Minuit
fitting routine provides also this matrix along with the fit
parameters \cite{MINUIT}. Denoting the set of fit parameters by $\xi$
and the Hessian error matrix by $H$, the fitting function $\chi^2$ can
be expanded around the minimum $\hat\xi$ as (See e.g. Ref.
\cite{Pumplin:2001ct}, here we follow the notation of Ref.
\cite{Hirai:2003pm})
\begin{equation}
  \Delta \chi^2 = \chi^2(\hat\xi+\delta\xi) - \chi^2(\hat\xi) 
    = \sum_{i,j} H_{ij} \delta\xi_i \delta\xi_j.
\end{equation}
The uncertainty of the fitted function $F(x,\hat\xi)$ is then
\begin{equation}
  [\delta F(x,\hat\xi)]^2 = \Delta\chi^2 
   \sum_{i,j} \left(\frac{\partial F(x,\hat\xi)}{\partial \xi_i}\right)
   H_{ij}^{-1}\left(\frac{\partial F(x,\hat\xi)}{\partial \xi_j}\right),
  \label{Errors}
\end{equation}
assuming linear error propagation.  However, the confidence region of
a multivariable fit is different than that of a single variable fit
and needs to be evaluated. The confidence level $P$ of the normal
distribution with $N$ degrees of freedom can be written as
\begin{equation}
  P = \int_0^{\Delta\chi^2} \frac{1}{2\,\Gamma(\frac{N}{2})} 
    \left(\frac{S}{2}\right)^{\frac{N}{2}-1} 
    \exp \left(\frac{-S}{2} \right)\,dS,
\end{equation}
where $\Gamma(n)$ is the Gamma function. For one-parameter fit the
one-$\sigma$ error range results confidence level $P=0.6826$ and
$\Delta\chi^2=1$. Requiring the same confidence level for $N$
parameters one can now calculate the $\Delta\chi^2$. For example, for
$N=16$ one obtains $\Delta\chi^2=18.11$.

\begin{figure}[htb]
\centering\includegraphics[width=\textwidth]{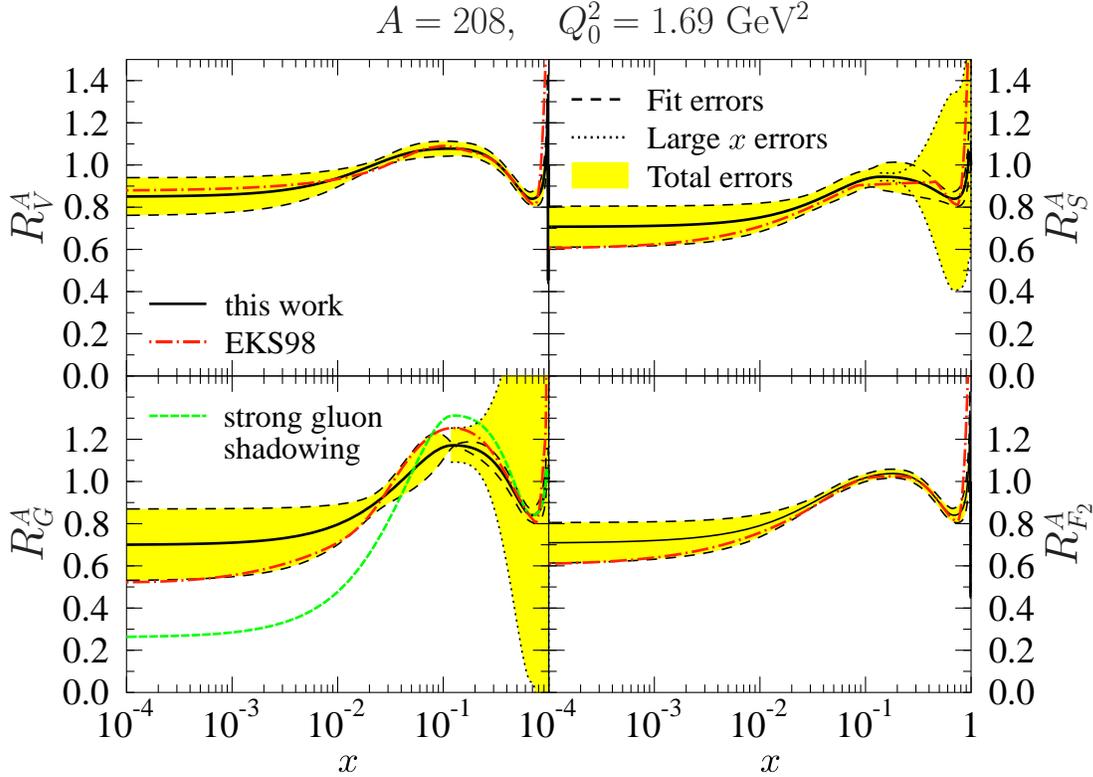}
\caption[]{\small (Colour online) Fit errors at the initial scale $Q_0^2=1.69$~GeV$^2$ for  Lead, 
  shown by the dashed lines.  For large-$x$ sea and gluon modifications the errors shown by the 
dotted      
  lines were calculated separately, see the text. 
  The shaded (yellow on-line) band is the total error estimate obtain, see the text.
  The corresponding {\em EKS98} results, evolved downwards from $Q_{0,EKS}^2=2.25$~GeV$^2$, are shown 
by the dot-dashed (red) lines.  An example of a stronger gluon shadowing is shown by dense-dashed 
(green) line.
  }
\label{Fig:Errors}
\end{figure}

The error limits obtained using this method
for the fit with the 16 free parameters in Table~\ref{Table:Params}
are shown by the dashed lines in Fig. \ref{Fig:Errors} for the ratios $R_V^A$, $R_S^A$, $R_G^A$ and 
$R_{F_2}^A$ in the case of a Lead nucleus, $A=208$. 
As can be seen from Fig. \ref{Fig:Errors}, and as expected on the basis of 
Sec.~\ref{sec:comparison_others}, the ratio $R_V^A$ is relatively well constrained.
Also $R_{F_2}^A$ is rather well under control. At large $x$ its errors naturally follow the small 
errors of $R_V^A$, thanks to the DIS data available.
In the small/mid-$x$ region both the DIS data and the DY data are necessary to pin down  
$R_V^A$ and $R_S^A$. Towards smaller values of $x$ the errors in $R_S^A$ get larger due to the lack 
of  
high-precision constraints for the sea quarks there, but are nevertheless still constrained.
As discussed in Sec.~\ref{sec:comparison_others}, the small-$x$ errors of $R_S^A$ shown, and thereby
those in $R_{F_2}^A$, are specific to the small-$x$ behaviour assumed. Hence, the error bars given 
here
are to be considered as lower limits.

For the gluons, the very-small-$x$ errors become quite large as there are no data constraints there 
to guide us. Similarly to the sea quark case, the error bars on gluon shadowing are fit function 
specific, and hence lower limits. However, as noticed in {\em EKS98} and originally in 
Ref.~\cite{Gousset:1996xt}  gluons do get somewhat better constrained at $x\sim 0.02-0.04$, thanks to 
the NMC data. Note that the zero-error we obtain at the peak of the gluon antishadowing bump is an 
artifact due to the interplay between the free parameters and the momentum sum rule. 

To get physically more relevant estimates on the sea quark and gluon uncertainties for the mid- and 
large-$x$ regions, we do the following. We free the parameters $y_a$, $p_{y_a}$, $y_e$, $p_{y_e}$ and 
$\beta$ (which control the magnitudes of the modifications in $R_S^A$ and $R_G^A$) while keeping the
location parameters $x_a$ and $x_e$ as well as the parameters controlling the small-$x$ behaviour 
fixed to the values quoted in Table~\ref{Table:Params}. Minimization of $\chi^2$ first with the freed 
sea quark parameters, then with the freed gluon parameters results in the wide bands shown by the 
dotted lines in Fig.~\ref{Fig:Errors}. This demonstrates clearly how badly the nuclear sea quark and 
gluon modifications are constrained in the large-$x$ region. 
Similar results have been presented before by the {\em HKN} group.
Thus, as the error estimates for the present analysis, we give the shaded (yellow on-line) bands of 
the small-$x$ and large-$x$ errors,  denoting them by "total errors" in Fig.~\ref{Fig:Errors}. 

In Fig. \ref{Fig:Errors} we also show the comparison with the {\em EKS98} modifications, evolved from 
a higher initial scale, $Q_{0,EKS}^2=2.25$ GeV$^2$, down to the present one, $Q_0^2=1.69$~GeV$^2$. 
Within the errors estimated, we can safely conclude that the old {\em EKS98} parametrization is fully 
consistent with the present $\chi^2$-minimization analysis. 
As discussed in the previous section, the fact that EKS98 sea quarks and gluons lie somewhat below 
the results from this work, is mainly due to the different functional forms assumed for the fit 
functions at small values of $x$.  We thus conclude that there is no need for releasing a new LO 
parametrization, since {\em EKS98} still works very well.

\section{Stronger gluon shadowing?}
\label{sec:strongershad}

Similarly to our earlier work {\em EKS98}, the present analysis suggests that the nuclear gluon 
modifications in the region $x\sim 0.02-0.04$ should be rather small, while the amounts 
of shadowing and thus antishadowing are much more weakly constrained. As the 
final task in this paper we discuss the possibility of a stronger gluon shadowing.
Our main motivation for doing this is the inclusive charged-hadron  data 
taken from D+Au collisions at RHIC by the 
BRAHMS collaboration \cite{Arsene:2004ux}, and the computation of the corresponding 
$p_T$ spectra in Ref.~\cite{Vogt:2004hf} using the strong gluon shadowing suggested 
in Refs.~\cite{Frankfurt:2003zd,Frankfurt:2002kd, Accardi:2004be}. These data are advocated
as a hint that a parton saturation regime could have been reached at RHIC \cite{satur}, so 
the degree of agreement with a DGLAP approach is of special interest.

We construct our strong gluon shadowing example by changing only the parameter $y_a$ 
for the Carbon reference nucleus in $R_G^A$. Then, as seen in Fig.~\ref{Fig:Errors} the changes in 
the region $x\sim 0.02-0.04$ remain small but the amounts of antishadowing and (through momentum 
conservation) shadowing change. Increasing $y_a$ from 1.071 to 1.2 deepens the saturation level of 
gluon shadowing in Lead considerably, from 0.7 to 0.26. At the same time, the goodness $\chi^2/N$ of 
the overall fit weakens only slightly, from 0.80 to 0.95, even if no $\chi^2$ minimization was 
performed. 

\begin{figure}[!]
\centering\includegraphics[width=15cm]{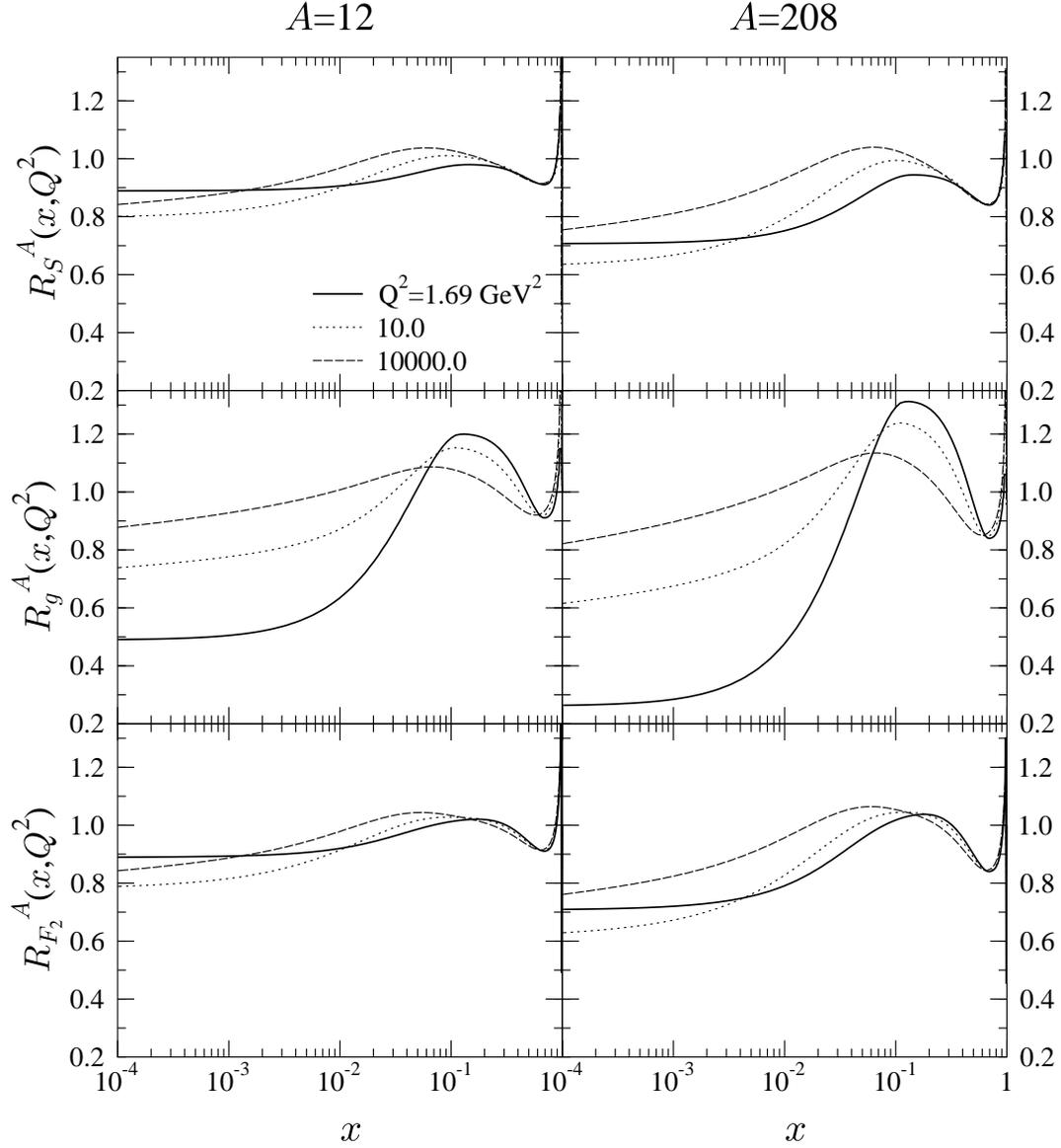}
\vspace{-0.5cm}
\caption[]{\small 
Scale evolution of the ratios $R_S^A$, $R_G^A$ and $R_{F_2}^A$ for Carbon and Lead 
in the case of the strong gluon shadowing example considered in Fig.~\ref{Fig:Errors}. Notice the 
initial negative $\log Q^2$ slopes of $R_S^A$ and hence also $R_{F_2}^A$ at small values of $x$. 
}
\label{Fig:sgs_scales}
\end{figure}

With the gluon shadowing much stronger than that of sea quarks,  the $\log Q^2$ slopes of $R_{F_2}^A$ 
at small $x$ are initially negative. At the same time, due to the stronger gluon antishadowing, the 
scale evolution of $R_S^A$ near $x\sim 0.1$ is slightly speeded up. These effects can be verified 
in Fig.~\ref{Fig:sgs_scales} (compare with Fig.~\ref{Fig:RAQ2}). In fact, the  latter effect is 
responsible for the deterioration of the goodness. We stress, however, that 
for this strong gluon shadowing example we have 
kept the quark sector as given in Table~\ref{Table:Params}. After minimization, the changes in  
$\chi^2/N$ would become even smaller, demonstrating the fact that quite large changes in the gluon 
sector induce only small changes in the global $\chi^2$.
This is interesting when compared with the results of de Sassot and Florian
\cite{deFlorian:2003qf}, who get considerably worse $\chi^2$ values
for stronger gluon shadowing. Apparently, the form of
their fit is such that stronger gluon shadowing in small-$x$ affects
in the region $x\sim 0.01-0.1$ as well, thus changing the fit there.

\begin{figure}[htb]
\vspace{-0.2cm}
\centering\includegraphics[width=14cm]{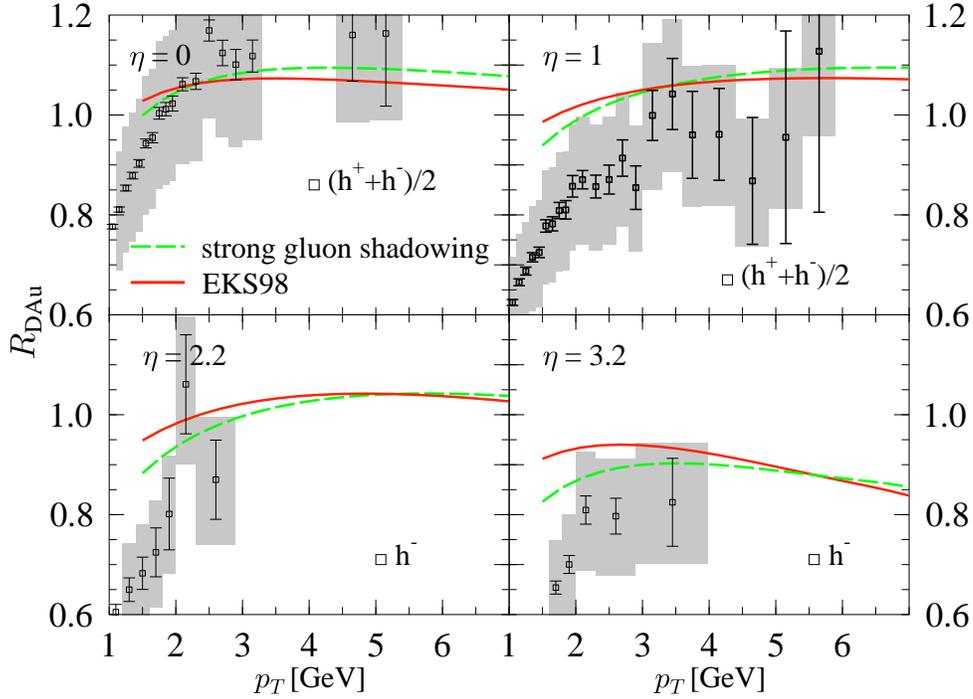}
\caption[]{\small (Colour online)
Minimum bias inclusive hadron production cross sections in d+Au collisions divided 
by that in p+p collisions at $\sqrt{s}_{NN}=200$ GeV at RHIC. The ratio  $R_{\mathrm{DAu}}$ is shown 
as a function of hadrons transverse momentum at four different pseudorapidities. 
The BRAHMS data \cite{Arsene:2004ux} are shown with the statistical error bars and the shaded 
systematic error limits.  A pQCD calculation for $h^++h^-$ production with the {\em EKS98} nuclear 
modifications and KKP fragmentation functions is shown by the solid lines (red) and that with the 
strong gluon shadowing by the dashed lines (green).
}
\label{Fig:Brahms}
\end{figure}

In Fig.~\ref{Fig:Brahms} we show the ratio $R_{DAu}$ for minimum bias single 
hadron production, defined as
\begin{equation}
R_{\mathrm{DAu}}= \frac
{\frac{1}{A}\frac{d \sigma^{\mathrm{DAu}}}{dp_T d\eta}}
{\frac{d \sigma^{\mathrm{pp}}}{dp_T d\eta}},
 \end{equation}
where $p_T$ and $\eta$ are the hadronic transverse momentum and pseudorapidity, correspondingly.
The BRAHMS data in the top panels are for $R_{\mathrm{DAu}}(h^+ + h^-)$ and in the bottom panels 
for $R_{\mathrm{DAu}}(h^-)$. The generic structure of the lowest order pQCD cross sections is given 
by 
\begin{equation}
\sigma^{AB\rightarrow h+X} = \sum_{ijkl} 
f_i^A (x_1,Q) \otimes f_j^B(x_2,Q) \otimes \sigma^{i+j\rightarrow k+l} 
\otimes D_{k\rightarrow h+X}(z,Q_f),
\label{Eq:Fragmentation}
\end{equation} 
where $h$ is the hadron type, $k$ labels the parton type,
$AB={\mathrm{DAu}}$, pp and $D_{k\rightarrow h+X}(z,Q_f)$ are the fragmentation functions
at a fractional energy $z=E_h/E_k$ and a factorization scale $Q_f$. Detailed formulation of the 
computation can be found e.g. in \cite{Eskola:2002kv}. Here we choose $Q$ as the transverse momentum 
of the parton and $Q_f$ as the transverse momentum of the hadron. 
We use the KKP fragmentation functions \cite{Kniehl:2000fe} and the CTEQ6L1 free proton PDFs. 
We do not make attempt to correct for the fact that the KKP fragmentation functions
correspond to the  average $h^++h^-$, even though the forward-rapidity data is for negative hadrons 
only.

At small pseudorapidities, where both quark and gluon-initiated processes are important, the stronger 
gluon antishadowing induces only a small correction to $R_{\mathrm{DAu}}$ but in a manner that the 
overall shape of the computed $R_{\mathrm{DAu}}$ agrees better with the BRAHMS data. At large 
pseudorapidities, corresponding to smaller $x_2$, gluons become dominant. As discussed in 
\cite{Eskola:2002kv},  hadron production at, say, 1.5 GeV is biased to partons at $p_T\sim 3$~GeV. 
Since $x_2=\frac{p_T}{\sqrt s}({\mathrm e}^{-\eta}+{\mathrm e}^{-y_2})$, small values of $x_2$ of the 
order 0.001, start to play a role at $\eta=3$. Integration over $y_2$ (or $x_2$) however, smears the 
effects of the nuclear modifications which is why we do not see a larger change in $R_{\mathrm{DAu}}$ 
with the stronger gluon shadowing example considered. As shown in Ref.~\cite{Vogt:2004hf}, even more 
dramatic small-$x$ behaviour of gluons, such as suggested in \cite{Frankfurt:2003zd,Frankfurt:2002kd, 
Accardi:2004be}, would obviously be needed to account for the BRAHMS data. 
Whether gluons with such shadowing, supplemented perhaps with stronger shadowing for the sea quarks 
as well, would maintain the good global fit to the DIS and DY data now obtained, remains to be seen. 
At the same time, dependence of the fragmentation functions on the hadron charge (negatives instead 
of 
the average $h^++h^-$), should be studied in more detail within a consistent DGLAP framework.

Due to the double integrations in computing the cross sections in Eq.~\ref{Eq:Fragmentation}, 
inclusion of the RHIC data for $R_{\mathrm{DAu}}$ in the global analysis is beyond the scope of the 
present paper. As further data constraints are absolutely necessary for pinning down the nuclear 
gluons, these data, in spite of their relatively large systematic errors, motivate us to do this in 
future.  

\section{Summary}
\label{sec:discussion}

In this study we have performed a global leading-order DGLAP analysis of the nPDFs using the {\em 
EKS98} framework introduced in \cite{Eskola:1998iy,Eskola:1998df}. Motivated by our previous work, we 
have introduced a piece-wize parametrization for the nuclear effects in the PDFs. Originally, the fit 
functions contained altogether 42 parameters. With the help of momentum and baryon number 
conservation and the experience from {\em EKS98}, we reduce the number of relevant fit parameters 
down to 16. 
A best fit to the nuclear DIS and DY data was searched for this set of parameters through automated  
minimization of $\chi^2$ using the Minuit program \cite{MINUIT}. As a result, a very good fit to the 
$N = 514$ data points at $Q^2\ge 1.69$~GeV$^2$ was found, giving $\chi^2/N=0.789$ (or 
$\chi^2/{\mathrm{d.o.f.}}=0.82$). 
No essential improvement over {\em EKS98} was found, however, as the {\em EKS98} modifications lead 
to an equally good fit quality, $\chi^2/N=0.809$ (for $N=479$ datapoints at 
$Q^2\ge Q^2_{0,{\mathrm{EKS98}}}$).

Relative to the old {\em EKS98}, the present analysis suggests slightly less shadowing for the gluons
and sea quarks. This, however, is merely due to the different forms of the fit functions adopted in 
the region where no stringent constraints from the data are available.
We also compared the obtained nuclear effects to those obtained by other global analyses, 
{\em HKM}, {\em HKN}, and {\em nDS}. The valence quark modifications do not deviate much from one set 
to another
but the smallest-$x$  and large-$x$ modifications of gluons and sea quarks differ in a major way.
This reflects the fact that especially the nuclear gluons are badly constrained in these regions.

To quantify the uncertainties in our analysis, we obtained the error estimates by using the Hessian 
method based on the information given by Minuit. The error estimates obtained also nicely further 
confirm the validity of {\em EKS98}, as it is shown to be fully consistent with the present analysis. 

To get a hold on the uncertainties in the large-$x$ regions of gluons and sea quarks, we computed the 
large-$x$ errors separately. These, considered together with the small-$x$ errors on the best fit 
confirm the conclusions from the comparison between different analyses: the valence quark 
distributions are relatively well, and independently from the fit-function form, 
constrained over the whole $x$ region. For the sea quarks, the large-$x$ ($x\gsim 0.3$) errors become 
very large, and for the small-$x$ behaviour clearly depends on the fit function form. For gluons, 
our analysis shows that presently one can to some extent constrain the gluons in the region 
$x\sim 0.02-0.04$ but hardly at all in the large-$x$ region, and only in a fit-function-dependent 
manner at small $x$ through momentum conservation. We also note that the relatively small error 
estimate obtained  at $x\sim 0.02-0.04$ for gluons may depend somewhat on the framework chosen, as 
the gluon fit parameters were drifting to the limits imposed. This obviously  leaves room for further 
improvements in the future. An obvious further improvement of the present analysis is its extension 
to NLO.

As the DIS and DY data are not able to stringently pin down the gluon modifications, 
further constraints are obviously needed. In thinking of possible additional data sets to be 
included in the global analysis in the future, we considered an example of a stronger gluon shadowing 
without doing a $\chi^2$ minimization. First, we showed that quite large variations in the gluon 
modifications can be absorbed in the quark sector and thus hidden by the good $\chi^2$ values 
obtained. Then, motivated by Ref.~\cite{Vogt:2004hf}, we computed the nuclear modification ratio 
$R_{\mathrm{DAu}}$ 
of inclusive hadron production in d+Au relative to that in pp,  using both
the {\em EKS98} modifications and the strong gluon shadowing example. Comparisons against the BRAHMS 
data \cite{Arsene:2004ux} here and in Ref.~\cite{Vogt:2004hf} lend support to  more shadowed gluons 
than in 
the present {\em EKS98} framework. At RHIC, the d+Au data is evidently very valuable for getting 
further constraints for nuclear gluons in particular. This in turn demonstrates the importance 
of running a parallel p+Pb program at the LHC, where pQCD factorization and nPDFs 
could be tested further in a wide range of $x$ and $Q^2$.

\section*{Acknowledgements}
CAS is supported by the 6th Framework Programme of the European Community under the Marie Curie 
contract MEIF-CT-2005-024624. We thank the Academy of Finland, Projects 73101, 80385, 206024 and 
115262 for financial support.


\begin{thebibliography}{99}

\bibitem{MRS03}
  A.~D.~Martin, R.~G.~Roberts, W.~J.~Stirling and R.~S.~Thorne,
  Eur.\ Phys.\ J.\  C {\bf 35} (2004) 325
  [arXiv:hep-ph/0308087].

\bibitem{CTEQ6}
J.~Pumplin, D.~R.~Stump, J.~Huston, H.~L.~Lai, P.~Nadolsky and W.~K.~Tung,
JHEP {\bf 0207} (2002) 012
[arXiv:hep-ph/0201195].

\bibitem{CTEQ61}
  D.~Stump, J.~Huston, J.~Pumplin, W.~K.~Tung, H.~L.~Lai, S.~Kuhlmann and J.~F.~Owens,
  JHEP {\bf 0310}, 046 (2003)
  [arXiv:hep-ph/0303013].

\bibitem{DGLAP} 
Y.~L.~Dokshitzer,
Perturbation Theory In Quantum 
Sov.\ Phys.\ JETP {\bf 46} (1977) 641
[Zh.\ Eksp.\ Teor.\ Fiz.\  {\bf 73} (1977) 1216];
V.~N.~Gribov and L.~N.~Lipatov,
Yad.\ Fiz.\  {\bf 15} (1972) 781
[Sov.\ J.\ Nucl.\ Phys.\  {\bf 15} (1972) 438];
V.~N.~Gribov and L.~N.~Lipatov,
Yad.\ Fiz.\  {\bf 15} (1972) 1218
[Sov.\ J.\ Nucl.\ Phys.\  {\bf 15} (1972) 675];
G.~Altarelli and G.~Parisi,
Nucl.\ Phys.\ B {\bf 126} (1977) 298.



  
\bibitem{Eskola:1998iy}
  K.~J.~Eskola, V.~J.~Kolhinen and P.~V.~Ruuskanen,
  Nucl.\ Phys.\ B {\bf 535} (1998) 351
  [arXiv:hep-ph/9802350].

\bibitem{Eskola:1998df}
  K.~J.~Eskola, V.~J.~Kolhinen and C.~A.~Salgado,
  Eur.\ Phys.\ J.\ C {\bf 9} (1999) 61
  [arXiv:hep-ph/9807297].

\bibitem{Hirai:2001np}
  M.~Hirai, S.~Kumano and M.~Miyama,
  Phys.\ Rev.\ D {\bf 64} (2001) 034003
  [arXiv:hep-ph/0103208].

\bibitem{Hirai:2004wq}
  M.~Hirai, S.~Kumano and T.~H.~Nagai,
  Phys.\ Rev.\ C {\bf 70} (2004) 044905
  [arXiv:hep-ph/0404093].

\bibitem{deFlorian:2003qf}
  D.~de Florian and R.~Sassot,
  Phys.\ Rev.\ D {\bf 69} (2004) 074028
  [arXiv:hep-ph/0311227].

\bibitem{Armesto:2006ph}
  N.~Armesto,
  J.\ Phys.\ G {\bf 32} (2006) R367
  [arXiv:hep-ph/0604108].
  
\bibitem{Arneodo:1992wf}
  M.~Arneodo,
  Phys.\ Rept.\  {\bf 240} (1994) 301.

   \bibitem{Accardi:2004be}
  A.~Accardi {\it et al.}, CERN Yellow Report for Hard Probes at the LHC, 
  ``Hard probes in heavy ion collisions at the LHC: PDFs, shadowing and p A
  collisions,''  arXiv:hep-ph/0308248; see Sec. 4.
 
\bibitem{Arneodo:1996ru}
  M.~Arneodo {\it et al.}  [New Muon Collaboration],
  Nucl.\ Phys.\ B {\bf 481} (1996) 23.

\bibitem{Qiu:1986wh}
  J.~w.~Qiu,
  Nucl.\ Phys.\ B {\bf 291} (1987) 746.

\bibitem{Frankfurt:1990xz}
  L.~L.~Frankfurt, M.~I.~Strikman and S.~Liuti,
  Phys.\ Rev.\ Lett.\  {\bf 65} (1990) 1725.


\bibitem{Eskola:1992zb}
  K.~J.~Eskola,
  Nucl.\ Phys.\ B {\bf 400} (1993) 240.

  \bibitem{Kumano:1992ef}
  S.~Kumano,
  Phys.\ Rev.\  C {\bf 48} (1993) 2016
  [arXiv:hep-ph/9303306].
  
  \bibitem{Kumano:1994pn}
  S.~Kumano,
  Phys.\ Rev.\  C {\bf 50} (1994) 1247
  [arXiv:hep-ph/9402321].
  
\bibitem{Indumathi:1996pb}
  D.~Indumathi and W.~Zhu,
  Z.\ Phys.\ C {\bf 74} (1997) 119
  [arXiv:hep-ph/9605417].

\bibitem{Indumathi:1996ky}
  D.~Indumathi,
  Z.\ Phys.\ C {\bf 76} (1997) 91
  [arXiv:hep-ph/9609361].

\bibitem{Frankfurt:2003zd}
  L.~Frankfurt, V.~Guzey and M.~Strikman,
  Phys.\ Rev.\ D {\bf 71} (2005) 054001
  [arXiv:hep-ph/0303022].

\bibitem{CTEQ_code} http://www.phys.psu.edu/\~{}cteq/


  \bibitem{Eskola:2006ux}
  K.~J.~Eskola and H.~Paukkunen,
  JHEP {\bf 0606} (2006) 008
  [arXiv:hep-ph/0603155].


\bibitem{Alde:1990im}
  D.~M.~Alde {\it et al.},
  Phys.\ Rev.\ Lett.\  {\bf 64} (1990) 2479.

\bibitem{Gomez:1993ri}
  J.~Gomez {\it et al.},
  Phys.\ Rev.\ D {\bf 49} (1994) 4348.


  
\bibitem{Adams:1995is}
  M.~R.~Adams {\it et al.}  [E665 Collaboration],
  Z.\ Phys.\ C {\bf 67} (1995) 403
  [arXiv:hep-ex/9505006].

  \bibitem{Amaudruz:1995tq}
  P.~Amaudruz {\it et al.}  [New Muon Collaboration],
  Nucl.\ Phys.\  B {\bf 441} (1995) 3
  [arXiv:hep-ph/9503291].
  
\bibitem{Arneodo:1995cs}
  M.~Arneodo {\it et al.}  [New Muon Collaboration.],
  Nucl.\ Phys.\ B {\bf 441} (1995) 12
  [arXiv:hep-ex/9504002].

\bibitem{Arneodo:1996rv}
  M.~Arneodo {\it et al.}  [New Muon Collaboration],
  Nucl.\ Phys.\ B {\bf 481} (1996) 3.

\bibitem{Vasilev:1999fa}
  M.~A.~Vasilev {\it et al.}  [FNAL E866 Collaboration],
  Phys.\ Rev.\ Lett.\  {\bf 83} (1999) 2304
  [arXiv:hep-ex/9906010].

\bibitem{HPC}
  Jan Czyzewski, K.J. Eskola and J. Qiu, at the III International Workshop on Hard Probes
  of Dense Matter, ECT., Trento, June 1995.
  
\bibitem{Eskola:2002us}
  K.~J.~Eskola, H.~Honkanen, V.~J.~Kolhinen and C.~A.~Salgado,
  Phys.\ Lett.\ B {\bf 532} (2002) 222
  [arXiv:hep-ph/0201256].

  
  \bibitem{Heli_thesis}
  H. Honkanen, PhD Thesis, Research Report 1/2005, University of Jyv\"askyl\"a, Department of 
Physics, January 2005; see Fig. 2.3.  
  
    
\bibitem{Qiu:2003vd}
  J.~w.~Qiu and I.~Vitev,
  Phys.\ Rev.\ Lett.\  {\bf 93}, 262301 (2004)
  [arXiv:hep-ph/0309094].

\bibitem{GLRMQ}
  L. V. Gribov, E. M. Levin and M. G. Ryskin, Phys. Rept. 100 (1983) 1;
  A. H. Mueller and J. Qiu, Nucl. Phys. B 268 (1986) 427.

\bibitem{Prytz:1993vr}
  K.~Prytz,
  Phys.\ Lett.\ B {\bf 311} (1993) 286.

\bibitem{MINUIT}
  F.~James, MINUIT Function Minimization and Error Analysis, 
  Reference Manual Version 94.1. CERN Program Library Long Writeup D506
  (Aug 1998).

\bibitem{Pumplin:2001ct}
  J.~Pumplin {\it et al.},
  Phys.\ Rev.\ D {\bf 65} (2002) 014013
  [arXiv:hep-ph/0101032].

\bibitem{Hirai:2003pm}
  M.~Hirai, S.~Kumano and N.~Saito  [Asymmetry Analysis Collaboration],
  Phys.\ Rev.\ D {\bf 69} (2004) 054021
  [arXiv:hep-ph/0312112].

  \bibitem{Gousset:1996xt}
  T.~Gousset and H.~J.~Pirner,
  Phys.\ Lett.\  B {\bf 375} (1996) 349
  [arXiv:hep-ph/9601242].
  
  \bibitem{Arsene:2004ux}
  I.~Arsene {\it et al.}  [BRAHMS Collaboration],
  Phys.\ Rev.\ Lett.\  {\bf 93} (2004) 242303
  [arXiv:nucl-ex/0403005].

  \bibitem{Vogt:2004hf}
  R.~Vogt,
  Phys.\ Rev.\  C {\bf 70} (2004) 064902.

 \bibitem{Frankfurt:2002kd}
  L.~Frankfurt, V.~Guzey, M.~McDermott and M.~Strikman,
  JHEP {\bf 0202} (2002) 027
  [arXiv:hep-ph/0201230].
  
  \bibitem{satur}
  R.~Baier, A.~Kovner and U.~A.~Wiedemann,
  Phys.\ Rev.\ D {\bf 68}, 054009 (2003);
  D.~Kharzeev, Y.~V.~Kovchegov and K.~Tuchin,
  Phys.\ Rev.\ D {\bf 68} (2003) 094013;
  J.~L.~Albacete, N.~Armesto, A.~Kovner, C.~A.~Salgado and U.~A.~Wiedemann,
  Phys.\ Rev.\ Lett.\  {\bf 92}, 082001 (2004).
  
  \bibitem{Eskola:2002kv}
  K.~J.~Eskola and H.~Honkanen,
  Nucl.\ Phys.\  A {\bf 713}, 167 (2003)
  [arXiv:hep-ph/0205048].
  
  \bibitem{Kniehl:2000fe}
  B.~A.~Kniehl, G.~Kramer and B.~Potter,
  Nucl.\ Phys.\  B {\bf 582} (2000) 514
  [arXiv:hep-ph/0010289].
  
\end{thebibliography}
\end{document}